\documentclass[11pt]{article}

\usepackage{amsmath,amssymb,amsthm}
\usepackage{xparse}
\usepackage[dvipsnames]{xcolor}
\usepackage{graphicx}
\usepackage{fullpage}
\usepackage{tikz}
\usepackage{bbold}
\usepackage{caption}
\usepackage{enumitem}

\usepackage[ruled,vlined,linesnumbered]{algorithm2e} 

\usepackage{euler} 
\usepackage{palatino} 
\usepackage{fourier-orns} 
\usepackage{pifont} 
\newcommand{\xmark}{\ding{55}}%

\usepackage{hyperref}
\hypersetup{
colorlinks=true,
breaklinks=true, 
urlcolor= OrangeRed, 
linkcolor= OrangeRed,
citecolor=OrangeRed,
}

\usepackage[colored]{shadethm} 
\definecolor{shadethmcolor}{rgb}{1,0.95,0.96} 
\definecolor{shaderulecolor}{rgb}{1,1,1} 

\newshadetheorem{theorem}{Theorem}[section] 
\newshadetheorem{env2}[theorem]{Lemma}
\newshadetheorem{env3}[theorem]{Proposition}

\NewDocumentEnvironment{lemma}{o}
  {\IfNoValueTF{#1}
      {\begin{env2}}
      {\begin{env2}[#1]}
  }
  {\end{env2}}
\NewDocumentEnvironment{proposition}{o}
  {\IfNoValueTF{#1}
      {\begin{env3}}
      {\begin{env3}[#1]}
  }
  {\end{env3}}

\DeclareMathOperator*{\treeNeST}{NeST}
\DeclareMathOperator*{\treeNEST}{NEST}
\DeclareMathOperator*{\LCA}{LCA}

\newcommand{\boundellipse}[3]
{(#1) ellipse (#2 and #3)
}

\newcommand{\Cross}{\mathbin{\tikz [x=1.4ex,y=1.4ex,line width=.2ex] \draw (0,0) -- (1,1) (0,1) -- (1,0);}}%
\newcommand{\BigCross}{\mathbin{\tikz [x=3ex,y=3ex,line width=.2ex] \draw (0,0) -- (1,1) (0,1) -- (1,0);}}%

\newcommand{\treeTA}{
\begin{tikzpicture}[xscale=0.2,yscale=0.2]
\tikzstyle{fleche}=[-,>=latex,thick]
\tikzstyle{noeud}=[fill=white,circle,draw]

\tikzstyle{texte}=[]

\tikzstyle{noeudn}=[fill=blue!40!red,circle,draw,scale=0.5]
\tikzstyle{noeudb}=[fill=blue!80!white,circle,draw,scale=0.5]
\tikzstyle{noeudr}=[fill=orange!85!white,circle,draw,scale=0.5]
\tikzstyle{noeudo}=[fill=white!10!red,circle,draw,scale=0.5]
\tikzstyle{noeudp}=[fill=red!20!pink,circle,draw,scale=0.5]
\tikzstyle{noeudg}=[fill=green!85!black!75!white,circle,draw,scale=0.5]

\tikzstyle{etiquette}=[midway]

\def\DistanceInterNiveaux{5}
\def\DistanceInterFeuilles{5}

\def\NiveauA{(-0)*\DistanceInterNiveaux}
\def\NiveauB{(-0.7)*\DistanceInterNiveaux}
\def\NiveauC{(-1.4)*\DistanceInterNiveaux}
\def\NiveauD{(-2.1)*\DistanceInterNiveaux}
\def\InterFeuilles{(0.5)*\DistanceInterFeuilles}

\node[texte] (T) at ({(0)*\InterFeuilles},{\NiveauA+2.5}) {\footnotesize $\tau_1$};
\node[noeudn] (A) at ({(0)*\InterFeuilles},{\NiveauA}) {};

\node[noeudb] (B1) at ({(-2)*\InterFeuilles},{\NiveauB}) {};
\node[noeudb] (B2) at ({(0)*\InterFeuilles},{\NiveauB}) {};
\node[noeudr] (B3) at ({(2)*\InterFeuilles},{\NiveauB}) {};

\node[noeudg] (C11) at ({(-2.5)*\InterFeuilles},{\NiveauC}) {};
\node[noeudo] (C12) at ({(-1.5)*\InterFeuilles},{\NiveauC}) {};
\node[noeudg] (C21) at ({(-0.5)*\InterFeuilles},{\NiveauC}) {};
\node[noeudo] (C22) at ({(0.5)*\InterFeuilles},{\NiveauC}) {};
\node[noeudp] (C31) at ({(2)*\InterFeuilles},{\NiveauC}) {};

\node[noeudo] (D111) at ({(-2.5)*\InterFeuilles},{\NiveauD}) {};
\node[noeudo] (D211) at ({(-0.5)*\InterFeuilles},{\NiveauD}) {};
\node[noeudo] (D311) at ({(1.6)*\InterFeuilles},{\NiveauD}) {};
\node[noeudo] (D312) at ({(2.4)*\InterFeuilles},{\NiveauD}) {};

\draw[fleche] (A)--(B1) node[etiquette] {};
\draw[fleche] (A)--(B2) node[etiquette] {};
\draw[fleche] (A)--(B3) node[etiquette] {};
\draw[fleche] (B1)--(C11) node[etiquette] {};
\draw[fleche] (B1)--(C12) node[etiquette] {};
\draw[fleche] (B2)--(C21) node[etiquette] {};
\draw[fleche] (B2)--(C22) node[etiquette] {};
\draw[fleche] (B3)--(C31) node[etiquette] {};
\draw[fleche] (C11)--(D111) node[etiquette] {};
\draw[fleche] (C21)--(D211) node[etiquette] {};
\draw[fleche] (C31)--(D311) node[etiquette] {};
\draw[fleche] (C31)--(D312) node[etiquette] {};

\tikzstyle{noeud}=[fill=blue!40!red,draw,scale=0.7]
\tikzstyle{noeudrouge}=[fill=orange!85!white,draw,scale=0.7]
\tikzstyle{noeudbleu}=[fill=blue!80!white,draw,scale=0.7]
\tikzstyle{noeudvert}=[fill=green!85!black!75!white,draw,scale=0.7]
\tikzstyle{noeudred}=[fill=white!10!red,draw,scale=0.7]
\tikzstyle{noeudpink}=[fill=red!20!pink,draw,scale=0.7]

\tikzstyle{etiquette}=[midway]

\def\shift{3}

\node[noeudred] (DAG6) at ({\shift+(4)*\InterFeuilles},{\NiveauD}) {};
\node[noeudpink] (DAG5) at ({\shift+(4.7)*\InterFeuilles},{\NiveauC}) {};
\node[noeudvert] (DAG4) at ({\shift+(3.3)*\InterFeuilles},{\NiveauC}) {};
\node[noeudrouge] (DAG3) at ({\shift+(4.7)*\InterFeuilles},{\NiveauB}) {};
\node[noeudbleu] (DAG2) at ({\shift+(3.3)*\InterFeuilles},{\NiveauB}) {};
\node[noeud] (DAG1) at ({\shift+(4)*\InterFeuilles},{\NiveauA}) {};

\draw[fleche] (DAG5)--(DAG6) node[etiquette] {\tiny $~~~~~2$};
\draw[fleche] (DAG4)--(DAG6) node[etiquette] {\tiny $1~~~~$};
\draw[fleche] (DAG3)--(DAG5) node[etiquette] {\tiny $~~~~1$};
\draw[fleche] (DAG2)--(DAG4) node[etiquette] {\tiny $1~~~$};
\draw[fleche] (DAG1)--(DAG3) node[etiquette] {\tiny $~~~~1$};
\draw[fleche] (DAG1)--(DAG2) node[etiquette] {\tiny $2~~~~~$};
\path[fleche] (DAG2) edge[bend left=-60] node {\tiny $1~~~$} (DAG6);
\end{tikzpicture}}

\newcommand{\treeTB}{
\begin{tikzpicture}[xscale=0.2,yscale=0.2]
\tikzstyle{fleche}=[-,>=latex,thick]
\tikzstyle{noeud}=[fill=white,circle,draw]
\tikzstyle{texte}=[]

\tikzstyle{noeudn}=[fill=blue!20!white,circle,draw,scale=0.5]
\tikzstyle{noeudb}=[fill=blue!80!white,circle,draw,scale=0.5]
\tikzstyle{noeudr}=[fill=red!85!white,circle,draw,scale=0.5]
\tikzstyle{noeudo}=[fill=white!10!red,circle,draw,scale=0.5]
\tikzstyle{noeudp}=[fill=red!20!pink,circle,draw,scale=0.5]
\tikzstyle{noeudg}=[fill=green!85!black!75!white,circle,draw,scale=0.5]

\tikzstyle{etiquette}=[midway]

\def\DistanceInterNiveaux{5}
\def\DistanceInterFeuilles{5}

\def\NiveauA{(-0)*\DistanceInterNiveaux}
\def\NiveauB{(-0.7)*\DistanceInterNiveaux}
\def\NiveauC{(-1.4)*\DistanceInterNiveaux}
\def\NiveauD{(-2.1)*\DistanceInterNiveaux}
\def\InterFeuilles{(0.5)*\DistanceInterFeuilles}

\node[texte] (T) at ({(0)*\InterFeuilles},{\NiveauA+2.5}) {\footnotesize $\tau_3$};
\node[noeudn] (A) at ({(0)*\InterFeuilles},{\NiveauA}) {};

\node[noeudb] (B1) at ({(-2)*\InterFeuilles},{\NiveauB}) {};
\node[noeudb] (B2) at ({(0)*\InterFeuilles},{\NiveauB}) {};
\node[noeudb] (B3) at ({(2)*\InterFeuilles},{\NiveauB}) {};

\node[noeudg] (C11) at ({(-2.5)*\InterFeuilles},{\NiveauC}) {};
\node[noeudo] (C12) at ({(-1.5)*\InterFeuilles},{\NiveauC}) {};
\node[noeudg] (C21) at ({(-0.5)*\InterFeuilles},{\NiveauC}) {};
\node[noeudo] (C22) at ({(0.5)*\InterFeuilles},{\NiveauC}) {};
\node[noeudg] (C31) at ({(1.5)*\InterFeuilles},{\NiveauC}) {};
\node[noeudo] (C32) at ({(2.5)*\InterFeuilles},{\NiveauC}) {};

\node[noeudo] (D111) at ({(-2.5)*\InterFeuilles},{\NiveauD}) {};
\node[noeudo] (D211) at ({(-0.5)*\InterFeuilles},{\NiveauD}) {};
\node[noeudo] (D311) at ({(1.5)*\InterFeuilles},{\NiveauD}) {};

\draw[fleche] (A)--(B1) node[etiquette] {};
\draw[fleche] (A)--(B2) node[etiquette] {};
\draw[fleche] (A)--(B3) node[etiquette] {};
\draw[fleche] (B1)--(C11) node[etiquette] {};
\draw[fleche] (B1)--(C12) node[etiquette] {};
\draw[fleche] (B2)--(C21) node[etiquette] {};
\draw[fleche] (B2)--(C22) node[etiquette] {};
\draw[fleche] (B3)--(C31) node[etiquette] {};
\draw[fleche] (B3)--(C32) node[etiquette] {};
\draw[fleche] (C11)--(D111) node[etiquette] {};
\draw[fleche] (C21)--(D211) node[etiquette] {};
\draw[fleche] (C31)--(D311) node[etiquette] {};

\tikzstyle{noeud}=[fill=blue!20!white,draw,scale=0.7]
\tikzstyle{noeudrouge}=[fill=red!85!white,draw,scale=0.7]
\tikzstyle{noeudbleu}=[fill=blue!80!white,draw,scale=0.7]
\tikzstyle{noeudvert}=[fill=green!85!black!75!white,draw,scale=0.7]
\tikzstyle{noeudred}=[fill=white!10!red,draw,scale=0.7]
\tikzstyle{noeudpink}=[fill=red!20!pink,draw,scale=0.7]

\tikzstyle{etiquette}=[midway]

\def\shift{3}

\node[noeudred] (DAG6) at ({\shift+(4)*\InterFeuilles},{\NiveauD}) {};
\node[noeudvert] (DAG4) at ({\shift+(4)*\InterFeuilles},{\NiveauC}) {};
\node[noeudbleu] (DAG2) at ({\shift+(4)*\InterFeuilles},{\NiveauB}) {};
\node[noeud] (DAG1) at ({\shift+(4)*\InterFeuilles},{\NiveauA}) {};

\draw[fleche] (DAG4)--(DAG6) node[etiquette] {\tiny $1~~~$};
\draw[fleche] (DAG2)--(DAG4) node[etiquette] {\tiny $1~~~$};
\draw[fleche] (DAG1)--(DAG2) node[etiquette] {\tiny $3~~~$};
\path[fleche] (DAG2) edge[bend left=-60] node {\tiny $1~~~$} (DAG6);
\end{tikzpicture}}

\newcommand{\treeTC}{
\begin{tikzpicture}[xscale=0.2,yscale=0.2]
\tikzstyle{fleche}=[-,>=latex,thick]
\tikzstyle{noeud}=[fill=white,circle,draw]
\tikzstyle{texte}=[]

\tikzstyle{noeudn}=[fill=black!30!green!70!blue,circle,draw,scale=0.5]
\tikzstyle{noeudb}=[fill=blue!80!white,circle,draw,scale=0.5]
\tikzstyle{noeudr}=[fill=red!20!yellow,circle,draw,scale=0.5]
\tikzstyle{noeudo}=[fill=white!10!red,circle,draw,scale=0.5]
\tikzstyle{noeudp}=[fill=red!20!pink,circle,draw,scale=0.5]
\tikzstyle{noeudg}=[fill=green!85!black!75!white,circle,draw,scale=0.5]
\tikzstyle{noeudpu}=[fill=purple!75!blue,circle,draw,scale=0.5]

\tikzstyle{etiquette}=[midway]

\def\DistanceInterNiveaux{5}
\def\DistanceInterFeuilles{5}

\def\NiveauA{(-0)*\DistanceInterNiveaux}
\def\NiveauB{(-0.7)*\DistanceInterNiveaux}
\def\NiveauC{(-1.4)*\DistanceInterNiveaux}
\def\NiveauD{(-2.1)*\DistanceInterNiveaux}
\def\InterFeuilles{(0.5)*\DistanceInterFeuilles}

\node[texte] (T) at ({(0)*\InterFeuilles},{\NiveauA+2.5}) {\footnotesize $\tau_2$};
\node[noeudn] (A) at ({(0)*\InterFeuilles},{\NiveauA}) {};

\node[noeudb] (B1) at ({(-2)*\InterFeuilles},{\NiveauB}) {};
\node[noeudr] (B2) at ({(0)*\InterFeuilles},{\NiveauB}) {};
\node[noeudpu] (B3) at ({(2)*\InterFeuilles},{\NiveauB}) {};

\node[noeudg] (C11) at ({(-2.5)*\InterFeuilles},{\NiveauC}) {};
\node[noeudo] (C12) at ({(-1.5)*\InterFeuilles},{\NiveauC}) {};
\node[noeudg] (C21) at ({(0)*\InterFeuilles},{\NiveauC}) {};
\node[noeudp] (C31) at ({(1.5)*\InterFeuilles},{\NiveauC}) {};
\node[noeudo] (C32) at ({(2.5)*\InterFeuilles},{\NiveauC}) {};

\node[noeudo] (D111) at ({(-2.5)*\InterFeuilles},{\NiveauD}) {};
\node[noeudo] (D211) at ({(0)*\InterFeuilles},{\NiveauD}) {};
\node[noeudo] (D311) at ({(1.1)*\InterFeuilles},{\NiveauD}) {};
\node[noeudo] (D312) at ({(1.9)*\InterFeuilles},{\NiveauD}) {};

\draw[fleche] (A)--(B1) node[etiquette] {};
\draw[fleche] (A)--(B2) node[etiquette] {};
\draw[fleche] (A)--(B3) node[etiquette] {};
\draw[fleche] (B1)--(C11) node[etiquette] {};
\draw[fleche] (B1)--(C12) node[etiquette] {};
\draw[fleche] (B2)--(C21) node[etiquette] {};
\draw[fleche] (B3)--(C31) node[etiquette] {};
\draw[fleche] (B3)--(C32) node[etiquette] {};
\draw[fleche] (C11)--(D111) node[etiquette] {};
\draw[fleche] (C21)--(D211) node[etiquette] {};
\draw[fleche] (C31)--(D311) node[etiquette] {};
\draw[fleche] (C31)--(D312) node[etiquette] {};

\tikzstyle{noeud}=[fill=black!30!green!70!blue,draw,scale=0.7]
\tikzstyle{noeudrouge}=[fill=red!20!yellow,draw,scale=0.7]
\tikzstyle{noeudbleu}=[fill=blue!80!white,draw,scale=0.7]
\tikzstyle{noeudvert}=[fill=green!85!black!75!white,draw,scale=0.7]
\tikzstyle{noeudred}=[fill=white!10!red,draw,scale=0.7]
\tikzstyle{noeudpink}=[fill=red!20!pink,draw,scale=0.7]
\tikzstyle{noeudbrown}=[fill=purple!75!blue,draw,scale=0.7]

\tikzstyle{etiquette}=[midway]

\def\shift{3}

\node[noeudred] (DAG6) at ({\shift+(4)*\InterFeuilles},{\NiveauD}) {};
\node[noeudpink] (DAG5) at ({\shift+(4.7)*\InterFeuilles},{\NiveauC}) {};
\node[noeudvert] (DAG4) at ({\shift+(3.3)*\InterFeuilles},{\NiveauC}) {};
\node[noeudbrown] (DAG33) at ({\shift+(5.2)*\InterFeuilles},{\NiveauB}) {};
\node[noeudrouge] (DAG3) at ({\shift+(4)*\InterFeuilles},{\NiveauB}) {};
\node[noeudbleu] (DAG2) at ({\shift+(2.8)*\InterFeuilles},{\NiveauB}) {};
\node[noeud] (DAG1) at ({\shift+(4)*\InterFeuilles},{\NiveauA}) {};

\draw[fleche] (DAG5)--(DAG6) node[etiquette] {\tiny $~~~~~2$};
\draw[fleche] (DAG4)--(DAG6) node[etiquette] {\tiny $1~~~~$};
\draw[fleche] (DAG33)--(DAG5) node[etiquette] {\tiny $~~~~1$};
\draw[fleche] (DAG3)--(DAG4) node[etiquette] {\tiny $~~~~~1$};
\draw[fleche] (DAG2)--(DAG4) node[etiquette] {\tiny $1~~~$};
\draw[fleche] (DAG1)--(DAG33) node[etiquette] {\tiny $~~~~~~1$};
\draw[fleche] (DAG1)--(DAG3) node[etiquette] {\tiny $~~~~1$};
\draw[fleche] (DAG1)--(DAG2) node[etiquette] {\tiny $1~~~~~~$};
\path[fleche] (DAG2) edge[bend left=-60] node {\tiny $1~~~~$} (DAG6);
\path[fleche] (DAG33) edge[bend left=60] node {\tiny $~~~~~1$} (DAG6);
\end{tikzpicture}}

\newcommand{\treeintro}{
\begin{tikzpicture}[xscale=0.3,yscale=0.6]
\tikzstyle{fleche}=[-,>=latex]
\tikzstyle{noeudblack}=[draw,circle,fill=black,scale=0.22]
\tikzstyle{texte}=[]
\def\localnodescalea{1}
\node[texte] (T) at ({0},{0+0.7}) {\footnotesize $\treeNeST(\tau)$};
\node[noeudblack,scale=\localnodescalea] (112299321720) at ({0},{0}) {};
\node[noeudblack,scale=\localnodescalea] (112299319424) at ({-5.5},{-1.0}) {};
\node[noeudblack,scale=\localnodescalea] (112299322728) at ({-4.5},{-1.0}) {};
\node[noeudblack,scale=\localnodescalea] (112299322896) at ({-3.5},{-1.0}) {};
\node[noeudblack,scale=\localnodescalea] (112299323176) at ({-3.5},{-1.8408964152537146}) {};
\node[noeudblack,scale=\localnodescalea] (112299322616) at ({-2.5},{-1.0}) {};
\node[noeudblack,scale=\localnodescalea] (112299323344) at ({-2.5},{-1.8408964152537146}) {};
\node[noeudblack,scale=\localnodescalea] (112299322672) at ({-1.5},{-1.0}) {};
\node[noeudblack,scale=\localnodescalea] (112299320264) at ({-1.5},{-1.8408964152537146}) {};
\node[noeudblack,scale=\localnodescalea] (112299322952) at ({-0.5},{-1.0}) {};
\node[noeudblack,scale=\localnodescalea] (112299322056) at ({-0.5},{-1.8408964152537146}) {};
\node[noeudblack,scale=\localnodescalea] (112299320656) at ({-0.5},{-2.600732100905307}) {};
\node[noeudblack,scale=\localnodescalea] (112299323120) at ({3.0},{-1.0}) {};
\node[noeudblack,scale=\localnodescalea] (112299321160) at ({0.5},{-1.8408964152537146}) {};
\node[noeudblack,scale=\localnodescalea] (112299320488) at ({1.5},{-1.8408964152537146}) {};
\node[noeudblack,scale=\localnodescalea] (112299323064) at ({1.5},{-2.600732100905307}) {};
\node[noeudblack,scale=\localnodescalea] (112299321776) at ({2.5},{-1.8408964152537146}) {};
\node[noeudblack,scale=\localnodescalea] (112299323232) at ({2.5},{-2.600732100905307}) {};
\node[noeudblack,scale=\localnodescalea] (112299323288) at ({4.5},{-1.8408964152537146}) {};
\node[noeudblack,scale=\localnodescalea] (112299320544) at ({3.5},{-2.600732100905307}) {};
\node[noeudblack,scale=\localnodescalea] (112299319984) at ({4.5},{-2.600732100905307}) {};
\node[noeudblack,scale=\localnodescalea] (112299319872) at ({5.5},{-2.600732100905307}) {};
\node[noeudblack,scale=\localnodescalea] (112299319760) at ({5.5},{-3.3078388820918545}) {};
\node[noeudblack,scale=\localnodescalea] (112299319536) at ({5.5},{-3.9765791870682765}) {};
\draw[fleche] (112299321720)--(112299319424) {};
\draw[fleche] (112299321720)--(112299322728) {};
\draw[fleche] (112299321720)--(112299322896) {};
\draw[fleche] (112299321720)--(112299322616) {};
\draw[fleche] (112299321720)--(112299322672) {};
\draw[fleche] (112299321720)--(112299322952) {};
\draw[fleche] (112299321720)--(112299323120) {};
\draw[fleche] (112299322896)--(112299323176) {};
\draw[fleche] (112299322616)--(112299323344) {};
\draw[fleche] (112299322672)--(112299320264) {};
\draw[fleche] (112299322952)--(112299322056) {};
\draw[fleche] (112299322056)--(112299320656) {};
\draw[fleche] (112299323120)--(112299321160) {};
\draw[fleche] (112299323120)--(112299320488) {};
\draw[fleche] (112299323120)--(112299321776) {};
\draw[fleche] (112299323120)--(112299323288) {};
\draw[fleche] (112299320488)--(112299323064) {};
\draw[fleche] (112299321776)--(112299323232) {};
\draw[fleche] (112299323288)--(112299320544) {};
\draw[fleche] (112299323288)--(112299319984) {};
\draw[fleche] (112299323288)--(112299319872) {};
\draw[fleche] (112299319872)--(112299319760) {};
\draw[fleche] (112299319760)--(112299319536) {};
\end{tikzpicture}\qquad
\begin{tikzpicture}[xscale=0.25,yscale=0.6]
\tikzstyle{fleche}=[-,>=latex]
\tikzstyle{noeudblack}=[draw,circle,fill=black,scale=0.22]
\tikzstyle{texte}=[]
\def\localnodescalea{1}
\node[texte] (T) at ({0},{0+0.7}) {\footnotesize $\tau$};
\node[noeudblack,scale=\localnodescalea] (112299322560) at ({0},{0}) {};
\node[noeudblack,scale=\localnodescalea] (112299322840) at ({-8.5},{-1.0}) {};
\node[noeudblack,scale=\localnodescalea] (112299319816) at ({-7.5},{-1.0}) {};
\node[noeudblack,scale=\localnodescalea] (112299320208) at ({-6.5},{-1.0}) {};
\node[noeudblack,scale=\localnodescalea] (112299319592) at ({-6.5},{-1.8408964152537146}) {};
\node[noeudblack,scale=\localnodescalea] (112299321440) at ({-5.0},{-1.0}) {};
\node[noeudblack,scale=\localnodescalea] (112299320992) at ({-5.5},{-1.8408964152537146}) {};
\node[noeudblack,scale=\localnodescalea] (112299320432) at ({-4.5},{-1.8408964152537146}) {};
\node[noeudblack,scale=\localnodescalea] (112299320040) at ({-3.5},{-1.0}) {};
\node[noeudblack,scale=\localnodescalea] (112299319704) at ({-3.5},{-1.8408964152537146}) {};
\node[noeudblack,scale=\localnodescalea] (112299320600) at ({-2.5},{-1.0}) {};
\node[noeudblack,scale=\localnodescalea] (112299321496) at ({-2.5},{-1.8408964152537146}) {};
\node[noeudblack,scale=\localnodescalea] (112299320376) at ({-2.5},{-2.600732100905307}) {};
\node[noeudblack,scale=\localnodescalea] (112299321608) at ({3.5},{-1.0}) {};
\node[noeudblack,scale=\localnodescalea] (112299321216) at ({-1.5},{-1.8408964152537146}) {};
\node[noeudblack,scale=\localnodescalea] (112299322112) at ({0.0},{-1.8408964152537146}) {};
\node[noeudblack,scale=\localnodescalea] (112299322448) at ({-0.5},{-2.600732100905307}) {};
\node[noeudblack,scale=\localnodescalea] (112299322392) at ({0.5},{-2.600732100905307}) {};
\node[noeudblack,scale=\localnodescalea] (112299320768) at ({2.0},{-1.8408964152537146}) {};
\node[noeudblack,scale=\localnodescalea] (112299321384) at ({1.5},{-2.600732100905307}) {};
\node[noeudblack,scale=\localnodescalea] (112299321272) at ({2.5},{-2.600732100905307}) {};
\node[noeudblack,scale=\localnodescalea] (112299321328) at ({6.0},{-1.8408964152537146}) {};
\node[noeudblack,scale=\localnodescalea] (112299319928) at ({3.5},{-2.600732100905307}) {};
\node[noeudblack,scale=\localnodescalea] (112299321552) at ({4.5},{-2.600732100905307}) {};
\node[noeudblack,scale=\localnodescalea] (112299322000) at ({7.0},{-2.600732100905307}) {};
\node[noeudblack,scale=\localnodescalea] (112299320712) at ({5.5},{-3.3078388820918545}) {};
\node[noeudblack,scale=\localnodescalea] (112299322168) at ({6.5},{-3.3078388820918545}) {};
\node[noeudblack,scale=\localnodescalea] (112299321944) at ({7.5},{-3.3078388820918545}) {};
\node[noeudblack,scale=\localnodescalea] (112299322336) at ({8.5},{-3.3078388820918545}) {};
\node[noeudblack,scale=\localnodescalea] (112299322784) at ({8.5},{-3.9765791870682765}) {};
\draw[fleche] (112299322560)--(112299322840) {};
\draw[fleche] (112299322560)--(112299319816) {};
\draw[fleche] (112299322560)--(112299320208) {};
\draw[fleche] (112299322560)--(112299321440) {};
\draw[fleche] (112299322560)--(112299320040) {};
\draw[fleche] (112299322560)--(112299320600) {};
\draw[fleche] (112299322560)--(112299321608) {};
\draw[fleche] (112299320208)--(112299319592) {};
\draw[fleche] (112299321440)--(112299320992) {};
\draw[fleche] (112299321440)--(112299320432) {};
\draw[fleche] (112299320040)--(112299319704) {};
\draw[fleche] (112299320600)--(112299321496) {};
\draw[fleche] (112299321496)--(112299320376) {};
\draw[fleche] (112299321608)--(112299321216) {};
\draw[fleche] (112299321608)--(112299322112) {};
\draw[fleche] (112299321608)--(112299320768) {};
\draw[fleche] (112299321608)--(112299321328) {};
\draw[fleche] (112299322112)--(112299322448) {};
\draw[fleche] (112299322112)--(112299322392) {};
\draw[fleche] (112299320768)--(112299321384) {};
\draw[fleche] (112299320768)--(112299321272) {};
\draw[fleche] (112299321328)--(112299319928) {};
\draw[fleche] (112299321328)--(112299321552) {};
\draw[fleche] (112299321328)--(112299322000) {};
\draw[fleche] (112299322000)--(112299320712) {};
\draw[fleche] (112299322000)--(112299322168) {};
\draw[fleche] (112299322000)--(112299321944) {};
\draw[fleche] (112299322000)--(112299322336) {};
\draw[fleche] (112299322336)--(112299322784) {};
\end{tikzpicture}\qquad
\begin{tikzpicture}[xscale=0.22,yscale=0.6]
\tikzstyle{fleche}=[-,>=latex]
\tikzstyle{noeudblack}=[draw,circle,fill=black,scale=0.22]
\tikzstyle{texte}=[]
\def\localnodescalea{1}
\node[texte] (T) at ({0},{0+0.7}) {\footnotesize $\treeNEST(\tau)$};
\node[noeudblack,scale=\localnodescalea] (112299323008) at ({0},{0}) {};
\node[noeudblack,scale=\localnodescalea] (112299320096) at ({-12.0},{-1.0}) {};
\node[noeudblack,scale=\localnodescalea] (112299319648) at ({-11.0},{-1.0}) {};
\node[noeudblack,scale=\localnodescalea] (112299319368) at ({-9.5},{-1.0}) {};
\node[noeudblack,scale=\localnodescalea] (112296115840) at ({-10.0},{-1.8408964152537146}) {};
\node[noeudblack,scale=\localnodescalea] (112290583272) at ({-9.0},{-1.8408964152537146}) {};
\node[noeudblack,scale=\localnodescalea] (112299319480) at ({-7.5},{-1.0}) {};
\node[noeudblack,scale=\localnodescalea] (112299182568) at ({-8.0},{-1.8408964152537146}) {};
\node[noeudblack,scale=\localnodescalea] (112292705504) at ({-7.0},{-1.8408964152537146}) {};
\node[noeudblack,scale=\localnodescalea] (112296195520) at ({-5.5},{-1.0}) {};
\node[noeudblack,scale=\localnodescalea] (112292705336) at ({-6.0},{-1.8408964152537146}) {};
\node[noeudblack,scale=\localnodescalea] (112292708136) at ({-5.0},{-1.8408964152537146}) {};
\node[noeudblack,scale=\localnodescalea] (112292705560) at ({-2.0},{-1.0}) {};
\node[noeudblack,scale=\localnodescalea] (112292708192) at ({-4.0},{-1.8408964152537146}) {};
\node[noeudblack,scale=\localnodescalea] (112292704328) at ({-3.0},{-1.8408964152537146}) {};
\node[noeudblack,scale=\localnodescalea] (112292708024) at ({-2.0},{-1.8408964152537146}) {};
\node[noeudblack,scale=\localnodescalea] (112292708248) at ({-0.5},{-1.8408964152537146}) {};
\node[noeudblack,scale=\localnodescalea] (112292704720) at ({-1.0},{-2.600732100905307}) {};
\node[noeudblack,scale=\localnodescalea] (112292705168) at ({0.0},{-2.600732100905307}) {};
\node[noeudblack,scale=\localnodescalea] (112292708080) at ({6.5},{-1.0}) {};
\node[noeudblack,scale=\localnodescalea] (112292707688) at ({1.0},{-1.8408964152537146}) {};
\node[noeudblack,scale=\localnodescalea] (112292708304) at ({2.5},{-1.8408964152537146}) {};
\node[noeudblack,scale=\localnodescalea] (112292707968) at ({2.0},{-2.600732100905307}) {};
\node[noeudblack,scale=\localnodescalea] (112292705280) at ({3.0},{-2.600732100905307}) {};
\node[noeudblack,scale=\localnodescalea] (112292705224) at ({4.5},{-1.8408964152537146}) {};
\node[noeudblack,scale=\localnodescalea] (112299355496) at ({4.0},{-2.600732100905307}) {};
\node[noeudblack,scale=\localnodescalea] (112299358360) at ({5.0},{-2.600732100905307}) {};
\node[noeudblack,scale=\localnodescalea] (112299354264) at ({9.0},{-1.8408964152537146}) {};
\node[noeudblack,scale=\localnodescalea] (112299359872) at ({6.0},{-2.600732100905307}) {};
\node[noeudblack,scale=\localnodescalea] (112299359760) at ({7.0},{-2.600732100905307}) {};
\node[noeudblack,scale=\localnodescalea] (112299359984) at ({10.0},{-2.600732100905307}) {};
\node[noeudblack,scale=\localnodescalea] (4546333328) at ({8.0},{-3.3078388820918545}) {};
\node[noeudblack,scale=\localnodescalea] (112279764832) at ({9.0},{-3.3078388820918545}) {};
\node[noeudblack,scale=\localnodescalea] (112277041448) at ({10.0},{-3.3078388820918545}) {};
\node[noeudblack,scale=\localnodescalea] (112299359648) at ({11.5},{-3.3078388820918545}) {};
\node[noeudblack,scale=\localnodescalea] (112299359424) at ({11.0},{-3.9765791870682765}) {};
\node[noeudblack,scale=\localnodescalea] (112299360096) at ({12.0},{-3.9765791870682765}) {};
\draw[fleche] (112299323008)--(112299320096) {};
\draw[fleche] (112299323008)--(112299319648) {};
\draw[fleche] (112299323008)--(112299319368) {};
\draw[fleche] (112299323008)--(112299319480) {};
\draw[fleche] (112299323008)--(112296195520) {};
\draw[fleche] (112299323008)--(112292705560) {};
\draw[fleche] (112299323008)--(112292708080) {};
\draw[fleche] (112299319368)--(112296115840) {};
\draw[fleche] (112299319368)--(112290583272) {};
\draw[fleche] (112299319480)--(112299182568) {};
\draw[fleche] (112299319480)--(112292705504) {};
\draw[fleche] (112296195520)--(112292705336) {};
\draw[fleche] (112296195520)--(112292708136) {};
\draw[fleche] (112292705560)--(112292708192) {};
\draw[fleche] (112292705560)--(112292704328) {};
\draw[fleche] (112292705560)--(112292708024) {};
\draw[fleche] (112292705560)--(112292708248) {};
\draw[fleche] (112292708248)--(112292704720) {};
\draw[fleche] (112292708248)--(112292705168) {};
\draw[fleche] (112292708080)--(112292707688) {};
\draw[fleche] (112292708080)--(112292708304) {};
\draw[fleche] (112292708080)--(112292705224) {};
\draw[fleche] (112292708080)--(112299354264) {};
\draw[fleche] (112292708304)--(112292707968) {};
\draw[fleche] (112292708304)--(112292705280) {};
\draw[fleche] (112292705224)--(112299355496) {};
\draw[fleche] (112292705224)--(112299358360) {};
\draw[fleche] (112299354264)--(112299359872) {};
\draw[fleche] (112299354264)--(112299359760) {};
\draw[fleche] (112299354264)--(112299359984) {};
\draw[fleche] (112299359984)--(4546333328) {};
\draw[fleche] (112299359984)--(112279764832) {};
\draw[fleche] (112299359984)--(112277041448) {};
\draw[fleche] (112299359984)--(112299359648) {};
\draw[fleche] (112299359648)--(112299359424) {};
\draw[fleche] (112299359648)--(112299360096) {};
\end{tikzpicture}
}

\newcommand{\treeTinsertleafA}{
\begin{tikzpicture}[xscale=0.15,yscale=0.15]
\tikzstyle{fleche}=[-,>=latex]
\tikzstyle{noeud}=[fill=white,circle,draw]

\tikzstyle{texte}=[]

\tikzstyle{noeudn}=[fill=black,circle,draw,scale=0.4]

\tikzstyle{etiquette}=[midway]

\def\DistanceInterNiveaux{5}
\def\DistanceInterFeuilles{5}

\def\NiveauA{(-0)*\DistanceInterNiveaux}
\def\NiveauB{(-0.7)*\DistanceInterNiveaux}
\def\NiveauC{(-1.4)*\DistanceInterNiveaux}
\def\NiveauD{(-2.1)*\DistanceInterNiveaux}

\def\InterFeuilles{(0.5)*\DistanceInterFeuilles}

\node[noeudn] (A) at ({(0)*\InterFeuilles},{\NiveauA}) {};

\node[noeudn] (B1) at ({(-2)*\InterFeuilles},{\NiveauB}) {};
\node[noeudn] (B2) at ({(0)*\InterFeuilles},{\NiveauB}) {};
\node[noeudn] (B3) at ({(2)*\InterFeuilles},{\NiveauB}) {};

\node[noeudn] (C11) at ({(-2.5)*\InterFeuilles},{\NiveauC}) {};
\node[noeudn] (C12) at ({(-1.5)*\InterFeuilles},{\NiveauC}) {};
\node[noeudn] (C21) at ({(-0.5)*\InterFeuilles},{\NiveauC}) {};
\node[noeudn] (C22) at ({(0.5)*\InterFeuilles},{\NiveauC}) {};
\node[noeudn] (C31) at ({(1.6)*\InterFeuilles},{\NiveauC}) {};
\node[noeudn] (C32) at ({(2.4)*\InterFeuilles},{\NiveauC}) {};

\node[noeudn] (D111) at ({(-2.5)*\InterFeuilles},{\NiveauD}) {};
\node[noeudn] (D211) at ({(-0.5)*\InterFeuilles},{\NiveauD}) {};

\tikzstyle{noeudb}=[fill=white,circle,scale=0.4]
\def\NiveauE{(-2.8)*\DistanceInterNiveaux}
\node[noeudb] (a) at ({(0)*\InterFeuilles},{\NiveauE}) {};

\draw[fleche] (A)--(B1) node[etiquette] {};
\draw[fleche] (A)--(B2) node[etiquette] {};
\draw[fleche] (A)--(B3) node[etiquette] {};
\draw[fleche] (B1)--(C11) node[etiquette] {};
\draw[fleche] (B1)--(C12) node[etiquette] {};
\draw[fleche] (B2)--(C21) node[etiquette] {};
\draw[fleche] (B2)--(C22) node[etiquette] {};
\draw[fleche] (B3)--(C31) node[etiquette] {};
\draw[fleche] (B3)--(C32) node[etiquette] {};
\draw[fleche] (C11)--(D111) node[etiquette] {};
\draw[fleche] (C21)--(D211) node[etiquette] {};

\end{tikzpicture}}

\newcommand{\treeTdelA}{
\begin{tikzpicture}[xscale=0.15,yscale=0.15]
\tikzstyle{fleche}=[-,>=latex]
\tikzstyle{noeud}=[fill=white,circle,draw]

\tikzstyle{texte}=[]

\tikzstyle{noeudn}=[fill=black,circle,draw,scale=0.4]
\tikzstyle{noeudR}=[fill=red,circle,draw,scale=0.4]

\tikzstyle{etiquette}=[midway]

\def\DistanceInterNiveaux{5}
\def\DistanceInterFeuilles{5}

\def\NiveauA{(-0)*\DistanceInterNiveaux}
\def\NiveauB{(-0.7)*\DistanceInterNiveaux}
\def\NiveauC{(-1.4)*\DistanceInterNiveaux}
\def\NiveauD{(-2.1)*\DistanceInterNiveaux}
\def\InterFeuilles{(0.5)*\DistanceInterFeuilles}

\node[noeudn] (A) at ({(0)*\InterFeuilles},{\NiveauA}) {};

\node[noeudn] (B1) at ({(-2)*\InterFeuilles},{\NiveauB}) {};
\node[noeudR] (B2) at ({(0)*\InterFeuilles},{\NiveauB}) {};
\node[noeudn] (B3) at ({(2)*\InterFeuilles},{\NiveauB}) {};

\node[noeudn] (C11) at ({(-2.5)*\InterFeuilles},{\NiveauC}) {};
\node[noeudn] (C12) at ({(-1.5)*\InterFeuilles},{\NiveauC}) {};
\node[noeudn] (C21) at ({(-0.5)*\InterFeuilles},{\NiveauC}) {};
\node[noeudn] (C22) at ({(0.5)*\InterFeuilles},{\NiveauC}) {};
\node[noeudn] (C31) at ({(1.6)*\InterFeuilles},{\NiveauC}) {};
\node[noeudn] (C32) at ({(2.4)*\InterFeuilles},{\NiveauC}) {};

\node[noeudn] (D111) at ({(-2.5)*\InterFeuilles},{\NiveauD}) {};
\node[noeudn] (D211) at ({(-0.5)*\InterFeuilles},{\NiveauD}) {};

\tikzstyle{noeudb}=[fill=white,circle,scale=0.4]
\def\NiveauE{(-2.8)*\DistanceInterNiveaux}
\node[noeudb] (a) at ({(0)*\InterFeuilles},{\NiveauE}) {};

\draw[fleche] (A)--(B1) node[etiquette] {};
\draw[fleche] (A)--(B2) node[etiquette] {};
\draw[fleche] (A)--(B3) node[etiquette] {};
\draw[fleche] (B1)--(C11) node[etiquette] {};
\draw[fleche] (B1)--(C12) node[etiquette] {};
\draw[fleche] (B2)--(C21) node[etiquette] {};
\draw[fleche] (B2)--(C22) node[etiquette] {};
\draw[fleche] (B3)--(C31) node[etiquette] {};
\draw[fleche] (B3)--(C32) node[etiquette] {};
\draw[fleche] (C11)--(D111) node[etiquette] {};
\draw[fleche] (C21)--(D211) node[etiquette] {};

\end{tikzpicture}}

\newcommand{\treeTdelB}{
\begin{tikzpicture}[xscale=0.15,yscale=0.15]
\tikzstyle{fleche}=[-,>=latex]
\tikzstyle{noeud}=[fill=white,circle,draw]

\tikzstyle{texte}=[]

\tikzstyle{noeudn}=[fill=black,circle,draw,scale=0.4]

\tikzstyle{etiquette}=[midway]

\def\DistanceInterNiveaux{5}
\def\DistanceInterFeuilles{5}

\def\NiveauA{(-0)*\DistanceInterNiveaux}
\def\NiveauB{(-0.7)*\DistanceInterNiveaux}
\def\NiveauC{(-1.4)*\DistanceInterNiveaux}
\def\NiveauD{(-2.1)*\DistanceInterNiveaux}
\def\InterFeuilles{(0.5)*\DistanceInterFeuilles}

\node[noeudn] (A) at ({(0)*\InterFeuilles},{\NiveauA}) {};

\node[noeudn] (B1) at ({(-2)*\InterFeuilles},{\NiveauB}) {};
\node[noeudn] (B3) at ({(2)*\InterFeuilles},{\NiveauB}) {};

\node[noeudn] (C11) at ({(-2.5)*\InterFeuilles},{\NiveauC}) {};
\node[noeudn] (C12) at ({(-1.5)*\InterFeuilles},{\NiveauC}) {};
\node[noeudn] (C21) at ({(-0.5)*\InterFeuilles},{\NiveauB}) {};
\node[noeudn] (C22) at ({(0.5)*\InterFeuilles},{\NiveauB}) {};
\node[noeudn] (C31) at ({(1.6)*\InterFeuilles},{\NiveauC}) {};
\node[noeudn] (C32) at ({(2.4)*\InterFeuilles},{\NiveauC}) {};

\node[noeudn] (D111) at ({(-2.5)*\InterFeuilles},{\NiveauD}) {};
\node[noeudn] (D211) at ({(-0.5)*\InterFeuilles},{\NiveauC}) {};

\tikzstyle{noeudb}=[fill=white,circle,scale=0.4]
\def\NiveauE{(-2.8)*\DistanceInterNiveaux}
\node[noeudb] (a) at ({(0)*\InterFeuilles},{\NiveauE}) {};

\draw[fleche] (A)--(B1) node[etiquette] {};
\draw[fleche] (A)--(B3) node[etiquette] {};
\draw[fleche] (B1)--(C11) node[etiquette] {};
\draw[fleche] (B1)--(C12) node[etiquette] {};
\draw[fleche] (A)--(C21) node[etiquette] {};
\draw[fleche] (A)--(C22) node[etiquette] {};
\draw[fleche] (B3)--(C31) node[etiquette] {};
\draw[fleche] (B3)--(C32) node[etiquette] {};
\draw[fleche] (C11)--(D111) node[etiquette] {};
\draw[fleche] (C21)--(D211) node[etiquette] {};

\end{tikzpicture}}

\newcommand{\treeTinsertleafC}{
\begin{tikzpicture}[xscale=0.15,yscale=0.15]
\tikzstyle{fleche}=[-,>=latex]
\tikzstyle{noeud}=[fill=white,circle,draw]

\tikzstyle{texte}=[]

\tikzstyle{noeudn}=[fill=black,circle,draw,scale=0.4]

\tikzstyle{flecheA}=[red,-,>=latex]
\tikzstyle{noeudA}=[fill=red,circle,draw,scale=0.4]

\tikzstyle{etiquette}=[midway]

\def\DistanceInterNiveaux{5}
\def\DistanceInterFeuilles{5}

\def\NiveauA{(-0)*\DistanceInterNiveaux}
\def\NiveauB{(-0.7)*\DistanceInterNiveaux}
\def\NiveauC{(-1.4)*\DistanceInterNiveaux}
\def\NiveauD{(-2.1)*\DistanceInterNiveaux}
\def\NiveauE{(-2.8)*\DistanceInterNiveaux}
\def\InterFeuilles{(0.5)*\DistanceInterFeuilles}

\node[noeudn] (A) at ({(0)*\InterFeuilles},{\NiveauA}) {};

\node[noeudn] (B1) at ({(-2)*\InterFeuilles},{\NiveauB}) {};
\node[noeudn] (B2) at ({(0)*\InterFeuilles},{\NiveauC}) {};

\node[noeudA] (NEW) at ({(1)*\InterFeuilles},{\NiveauB}) {};

\node[noeudn] (C11) at ({(-2.5)*\InterFeuilles},{\NiveauC}) {};
\node[noeudn] (C12) at ({(-1.5)*\InterFeuilles},{\NiveauC}) {};
\node[noeudn] (C21) at ({(-0.5)*\InterFeuilles},{\NiveauD}) {};
\node[noeudn] (C22) at ({(0.5)*\InterFeuilles},{\NiveauD}) {};

\node[noeudn] (C31) at ({(2)*\InterFeuilles},{\NiveauC}) {};

\node[noeudn] (D111) at ({(-2.5)*\InterFeuilles},{\NiveauD}) {};
\node[noeudn] (D211) at ({(-0.5)*\InterFeuilles},{\NiveauE}) {};

\node[noeudn] (D311) at ({(1.6)*\InterFeuilles},{\NiveauD}) {};
\node[noeudn] (D312) at ({(2.4)*\InterFeuilles},{\NiveauD}) {};

\draw[fleche] (A)--(B1) node[etiquette] {};
\draw[fleche] (A)--(NEW) node[etiquette] {};
\draw[fleche] (B1)--(C11) node[etiquette] {};
\draw[fleche] (B1)--(C12) node[etiquette] {};

\draw[fleche] (NEW)--(B2) node[etiquette] {};

\draw[fleche] (B2)--(C21) node[etiquette] {};
\draw[fleche] (B2)--(C22) node[etiquette] {};
\draw[fleche] (NEW)--(C31) node[etiquette] {};
\draw[fleche] (C11)--(D111) node[etiquette] {};
\draw[fleche] (C21)--(D211) node[etiquette] {};
\draw[fleche] (C31)--(D311) node[etiquette] {};
\draw[fleche] (C31)--(D312) node[etiquette] {};
\end{tikzpicture}}

\newcommand{\mappingA}{
\begin{tikzpicture}[xscale=0.2,yscale=0.2]
\tikzstyle{fleche}=[-,>=latex]
\tikzstyle{flecher}=[->,>=latex,red]
\tikzstyle{noeud}=[fill=white,circle,draw]

\tikzstyle{texte}=[]

\tikzstyle{noeudn}=[fill=black,circle,draw,scale=0.4]
\tikzstyle{noeudR}=[fill=red,circle,draw,scale=0.4]

\tikzstyle{etiquette}=[midway]

\def\DistanceInterNiveaux{5}
\def\DistanceInterFeuilles{5}

\def\NiveauA{(-0)*\DistanceInterNiveaux}
\def\NiveauB{(-0.7)*\DistanceInterNiveaux}
\def\NiveauC{(-1.4)*\DistanceInterNiveaux}
\def\NiveauD{(-2.1)*\DistanceInterNiveaux}
\def\InterFeuilles{(0.5)*\DistanceInterFeuilles}

\node[noeudn] (A) at ({(0)*\InterFeuilles},{\NiveauA}) {};
\node[texte] (TAA) at ({(0)*\InterFeuilles},{\NiveauA+2}) {\small$\tau$};

\node[texte] (TA) at ({(0)*\InterFeuilles-5},{\NiveauA}) {\tiny$\text{LCA}(v_1,v_2)$};

\node[noeudn] (B1) at ({(-2)*\InterFeuilles},{\NiveauB}) {};
\node[texte] (TB1) at ({(-2.5)*\InterFeuilles},{\NiveauB}) {\tiny$v_3$};

\node[noeudn] (B2) at ({(0)*\InterFeuilles},{\NiveauB}) {};
\node[texte] (TB2) at ({(-0.5)*\InterFeuilles},{\NiveauB}) {\tiny$v_1$};

\node[noeudn] (B3) at ({(2)*\InterFeuilles},{\NiveauB}) {};
\node[texte] (TB3) at ({(1.5)*\InterFeuilles},{\NiveauB}) {\tiny$v_2$};

\node[noeudn] (C11) at ({(-2.5)*\InterFeuilles},{\NiveauC}) {};
\node[noeudn] (C12) at ({(-1.5)*\InterFeuilles},{\NiveauC}) {};
\node[noeudn] (C21) at ({(-0.5)*\InterFeuilles},{\NiveauC}) {};
\node[noeudn] (C22) at ({(0.5)*\InterFeuilles},{\NiveauC}) {};
\node[noeudn] (C31) at ({(1.6)*\InterFeuilles},{\NiveauC}) {};
\node[noeudn] (C32) at ({(2.4)*\InterFeuilles},{\NiveauC}) {};

\node[noeudn] (D111) at ({(-2.5)*\InterFeuilles},{\NiveauD}) {};
\node[noeudn] (D211) at ({(-0.5)*\InterFeuilles},{\NiveauD}) {};

\tikzstyle{noeudb}=[fill=white,circle,scale=0.4]
\def\NiveauE{(-2.8)*\DistanceInterNiveaux}
\node[noeudb] (a) at ({(0)*\InterFeuilles},{\NiveauE}) {};

\draw[fleche] (A)--(B1) node[etiquette] {};
\draw[fleche] (A)--(B2) node[etiquette] {};
\draw[fleche] (A)--(B3) node[etiquette] {};
\draw[fleche] (B1)--(C11) node[etiquette] {};
\draw[fleche] (B1)--(C12) node[etiquette] {};
\draw[fleche] (B2)--(C21) node[etiquette] {};
\draw[fleche] (B2)--(C22) node[etiquette] {};
\draw[fleche] (B3)--(C31) node[etiquette] {};
\draw[fleche] (B3)--(C32) node[etiquette] {};
\draw[fleche] (C11)--(D111) node[etiquette] {};
\draw[fleche] (C21)--(D211) node[etiquette] {};

\def\Shift{(25)}

\node[noeudn] (2A) at ({\Shift+(0)*\InterFeuilles},{\NiveauA}) {};
\node[texte] (T2A) at ({\Shift+(0)*\InterFeuilles},{\NiveauA+2}) {\small$\theta$};

\node[noeudn] (2B1) at ({\Shift+(-2)*\InterFeuilles},{\NiveauB}) {};
\node[texte] (T2B1) at ({\Shift+(-1.3)*\InterFeuilles},{\NiveauB}) {\tiny$w_3$};

\node[noeudn] (2B2) at ({\Shift+(0)*\InterFeuilles},{\NiveauC}) {};
\node[texte] (T2B2) at ({\Shift+(0.7)*\InterFeuilles},{\NiveauC}) {\tiny$w_1$};

\node[noeudR] (2NEW) at ({\Shift+(1)*\InterFeuilles},{\NiveauB}) {};
\node[texte] (T2NEW) at ({\Shift+(1)*\InterFeuilles+5},{\NiveauB}) {\tiny$\text{LCA}(w_1,w_2)$};

\node[noeudn] (2C11) at ({\Shift+(-2.5)*\InterFeuilles},{\NiveauC}) {};
\node[noeudn] (2C12) at ({\Shift+(-1.5)*\InterFeuilles},{\NiveauC}) {};
\node[noeudn] (2C21) at ({\Shift+(-0.5)*\InterFeuilles},{\NiveauD}) {};
\node[noeudn] (2C22) at ({\Shift+(0.5)*\InterFeuilles},{\NiveauD}) {};

\node[noeudn] (2C31) at ({\Shift+(2)*\InterFeuilles},{\NiveauC}) {};
\node[texte] (T2C31) at ({\Shift+(2.7)*\InterFeuilles},{\NiveauC}) {\tiny$w_2$};

\node[noeudn] (2D111) at ({\Shift+(-2.5)*\InterFeuilles},{\NiveauD}) {};
\node[noeudn] (2D211) at ({\Shift+(-0.5)*\InterFeuilles},{\NiveauE}) {};

\node[noeudn] (2D311) at ({\Shift+(1.6)*\InterFeuilles},{\NiveauD}) {};
\node[noeudn] (2D312) at ({\Shift+(2.4)*\InterFeuilles},{\NiveauD}) {};

\draw[fleche] (2A)--(2B1) node[etiquette] {};
\draw[fleche] (2A)--(2NEW) node[etiquette] {};
\draw[fleche] (2B1)--(2C11) node[etiquette] {};
\draw[fleche] (2B1)--(2C12) node[etiquette] {};
\draw[fleche] (2NEW)--(2B2) node[etiquette] {};
\draw[fleche] (2B2)--(2C21) node[etiquette] {};
\draw[fleche] (2B2)--(2C22) node[etiquette] {};
\draw[fleche] (2NEW)--(2C31) node[etiquette] {};
\draw[fleche] (2C11)--(2D111) node[etiquette] {};
\draw[fleche] (2C21)--(2D211) node[etiquette] {};
\draw[fleche] (2C31)--(2D311) node[etiquette] {};
\draw[fleche] (2C31)--(2D312) node[etiquette] {};

\draw[flecher] (B1) edge[bend left=30] (2B1) node[etiquette] {};

\draw[flecher] (B2) edge[bend left=-30] (2B2) node[etiquette] {};

\draw[flecher] (B3) edge[bend left=10] (2C31) node[etiquette] {};

\end{tikzpicture}}

\newcommand{\treeonechildallchildrenA}{
\begin{tikzpicture}[xscale=0.15,yscale=0.15]
\tikzstyle{fleche}=[-,>=latex]
\tikzstyle{noeud}=[fill=white,circle,draw]
\tikzstyle{noeudn}=[fill=black,circle,draw,scale=0.4]
\tikzstyle{noeudb}=[fill=white,circle,scale=0.4]

\tikzstyle{texte}=[]

\tikzstyle{etiquette}=[midway]

\def\DistanceInterNiveaux{5}
\def\DistanceInterFeuilles{5}

\def\NiveauA{(-0)*\DistanceInterNiveaux}
\def\NiveauB{(-0.7)*\DistanceInterNiveaux}
\def\NiveauC{(-1.4)*\DistanceInterNiveaux}
\def\NiveauD{(-2.1)*\DistanceInterNiveaux}

\def\InterFeuilles{(0.5)*\DistanceInterFeuilles}

\node[noeudn] (A) at ({(0)*\InterFeuilles},{\NiveauA}) {};
\node[texte] (TA) at ({(0)*\InterFeuilles+(-1.5)},{\NiveauA}) {\tiny$v$};

\node[noeudn] (B1) at ({(-2)*\InterFeuilles},{\NiveauB}) {};
\node[texte] (TB1) at ({(-2)*\InterFeuilles+(-1.5)},{\NiveauB}) {\tiny$w$};

\node[noeudn] (B2) at ({(0)*\InterFeuilles},{\NiveauB}) {};
\node[noeudn] (B3) at ({(2)*\InterFeuilles},{\NiveauB}) {};

\node[noeudn] (C11) at ({(-3)*\InterFeuilles},{\NiveauC}) {};
\node[noeudn] (C12) at ({(-2)*\InterFeuilles},{\NiveauC}) {};
\node[noeudn] (C13) at ({(-1)*\InterFeuilles},{\NiveauC}) {};

\node[noeudb] (D000) at ({(-3)*\InterFeuilles},{\NiveauD}) {};

\draw[fleche] (A)--(B1) node[etiquette] {};
\draw[fleche] (A)--(B2) node[etiquette] {};
\draw[fleche] (A)--(B3) node[etiquette] {};
\draw[fleche] (B1)--(C11) node[etiquette] {};
\draw[fleche] (B1)--(C12) node[etiquette] {};
\draw[fleche] (B1)--(C13) node[etiquette] {};

\end{tikzpicture}
}

\newcommand{\treeonechildallchildrenB}{
\begin{tikzpicture}[xscale=0.15,yscale=0.15]
\tikzstyle{fleche}=[-,>=latex]
\tikzstyle{noeud}=[fill=white,circle,draw]
\tikzstyle{noeudn}=[fill=black,circle,draw,scale=0.4]
\tikzstyle{noeudb}=[fill=white,circle,scale=0.4]

\tikzstyle{texte}=[]

\tikzstyle{etiquette}=[midway]

\def\DistanceInterNiveaux{5}
\def\DistanceInterFeuilles{5}

\def\NiveauA{(-0)*\DistanceInterNiveaux}
\def\NiveauB{(-0.7)*\DistanceInterNiveaux}
\def\NiveauC{(-1.4)*\DistanceInterNiveaux}
\def\NiveauD{(-2.1)*\DistanceInterNiveaux}

\def\InterFeuilles{(0.5)*\DistanceInterFeuilles}

\node[noeudn] (A) at ({(0)*\InterFeuilles},{\NiveauA}) {};
\node[texte] (TA) at ({(0)*\InterFeuilles+(-1.5)},{\NiveauA}) {\tiny$v$};

\node[noeudn] (B1) at ({(-2)*\InterFeuilles},{\NiveauB}) {};
\node[texte] (TB1) at ({(-2)*\InterFeuilles+(-1.5)},{\NiveauB}) {\tiny$w$};

\node[noeudn] (B2) at ({(0)*\InterFeuilles},{\NiveauB}) {};
\node[noeudn] (B3) at ({(2)*\InterFeuilles},{\NiveauB}) {};

\node[noeudn] (C11) at ({(-3)*\InterFeuilles},{\NiveauC}) {};
\node[noeudn] (C12) at ({(-2)*\InterFeuilles},{\NiveauC}) {};
\node[noeudn] (C13) at ({(-1)*\InterFeuilles},{\NiveauC}) {};

\node[noeudb] (D000) at ({(-3)*\InterFeuilles},{\NiveauD}) {};

\draw[fleche] (A)--(B1) node[etiquette] {};
\draw[fleche] (A)--(B2) node[etiquette] {};
\draw[fleche] (A)--(B3) node[etiquette] {};
\draw[fleche] (B1)--(C11) node[etiquette] {};
\draw[fleche] (B1)--(C12) node[etiquette] {};
\draw[fleche] (B1)--(C13) node[etiquette] {};

\draw[red,thick] \boundellipse{-5,-5.5}{2.6}{0.5};

\end{tikzpicture}
}

\newcommand{\treeonechildallchildrenC}{
\begin{tikzpicture}[xscale=0.15,yscale=0.15]
\tikzstyle{fleche}=[-,>=latex]
\tikzstyle{noeud}=[fill=white,circle,draw]
\tikzstyle{noeudn}=[fill=black,circle,draw,scale=0.4]
\tikzstyle{noeudb}=[fill=white,circle,scale=0.4]
\tikzstyle{noeudr}=[fill=red,circle,draw,scale=0.4]

\tikzstyle{texte}=[]

\tikzstyle{etiquette}=[midway]

\def\DistanceInterNiveaux{5}
\def\DistanceInterFeuilles{5}

\def\NiveauA{(-0)*\DistanceInterNiveaux}
\def\NiveauB{(-0.7)*\DistanceInterNiveaux}
\def\NiveauC{(-1.4)*\DistanceInterNiveaux}
\def\NiveauD{(-2.1)*\DistanceInterNiveaux}

\def\InterFeuilles{(0.5)*\DistanceInterFeuilles}

\node[noeudn] (A) at ({(0)*\InterFeuilles},{\NiveauA}) {};
\node[texte] (TA) at ({(0)*\InterFeuilles+(-1.5)},{\NiveauA}) {\tiny$v$};

\node[noeudn] (B1) at ({(-2)*\InterFeuilles},{\NiveauB}) {};
\node[texte] (TB1) at ({(-2)*\InterFeuilles+(-1.5)},{\NiveauB}) {\tiny$w$};

\node[noeudn] (B2) at ({(0)*\InterFeuilles},{\NiveauB}) {};
\node[noeudn] (B3) at ({(2)*\InterFeuilles},{\NiveauB}) {};

\node[noeudr] (NEW) at ({(-2)*\InterFeuilles},{\NiveauC}) {};

\node[noeudn] (C11) at ({(-3)*\InterFeuilles},{\NiveauD}) {};
\node[noeudn] (C12) at ({(-2)*\InterFeuilles},{\NiveauD}) {};
\node[noeudn] (C13) at ({(-1)*\InterFeuilles},{\NiveauD}) {};

\draw[fleche] (A)--(B1) node[etiquette] {};
\draw[fleche] (A)--(B2) node[etiquette] {};
\draw[fleche] (A)--(B3) node[etiquette] {};

\draw[fleche] (B1)--(NEW) node[etiquette] {};

\draw[fleche] (NEW)--(C11) node[etiquette] {};
\draw[fleche] (NEW)--(C12) node[etiquette] {};
\draw[fleche] (NEW)--(C13) node[etiquette] {};

\end{tikzpicture}
}

\newcommand{\treeonechildallchildrenD}{
\begin{tikzpicture}[xscale=0.15,yscale=0.15]
\tikzstyle{fleche}=[-,>=latex]
\tikzstyle{noeud}=[fill=white,circle,draw]
\tikzstyle{noeudn}=[fill=black,circle,draw,scale=0.4]
\tikzstyle{noeudb}=[fill=white,circle,scale=0.4]

\tikzstyle{texte}=[]

\tikzstyle{etiquette}=[midway]

\def\DistanceInterNiveaux{5}
\def\DistanceInterFeuilles{5}

\def\NiveauA{(-0)*\DistanceInterNiveaux}
\def\NiveauB{(-0.7)*\DistanceInterNiveaux}
\def\NiveauC{(-1.4)*\DistanceInterNiveaux}
\def\NiveauD{(-2.1)*\DistanceInterNiveaux}

\def\InterFeuilles{(0.5)*\DistanceInterFeuilles}

\node[noeudn] (A) at ({(0)*\InterFeuilles},{\NiveauA}) {};
\node[texte] (TA) at ({(0)*\InterFeuilles+(-1.5)},{\NiveauA}) {\tiny$v$};

\node[noeudn] (B1) at ({(-2)*\InterFeuilles},{\NiveauB}) {};
\node[texte] (TB1) at ({(-2)*\InterFeuilles+(-1.5)},{\NiveauB}) {\tiny$w$};

\node[noeudn] (B2) at ({(0)*\InterFeuilles},{\NiveauB}) {};
\node[noeudn] (B3) at ({(2)*\InterFeuilles},{\NiveauB}) {};

\node[noeudn] (C11) at ({(-3)*\InterFeuilles},{\NiveauC}) {};
\node[noeudn] (C12) at ({(-2)*\InterFeuilles},{\NiveauC}) {};
\node[noeudn] (C13) at ({(-1)*\InterFeuilles},{\NiveauC}) {};

\node[noeudb] (D000) at ({(-3)*\InterFeuilles},{\NiveauD}) {};

\draw[fleche] (A)--(B1) node[etiquette] {};
\draw[fleche] (A)--(B2) node[etiquette] {};
\draw[fleche] (A)--(B3) node[etiquette] {};
\draw[fleche] (B1)--(C11) node[etiquette] {};
\draw[fleche] (B1)--(C12) node[etiquette] {};
\draw[fleche] (B1)--(C13) node[etiquette] {};

\draw[red,thick] \boundellipse{-2.5,-1.7}{0.5}{0.5};

\end{tikzpicture}
}

\newcommand{\treeonechildallchildrenE}{
\begin{tikzpicture}[xscale=0.15,yscale=0.15]
\tikzstyle{fleche}=[-,>=latex]
\tikzstyle{noeud}=[fill=white,circle,draw]
\tikzstyle{noeudn}=[fill=black,circle,draw,scale=0.4]
\tikzstyle{noeudb}=[fill=white,circle,scale=0.4]
\tikzstyle{noeudr}=[fill=red,circle,draw,scale=0.4]

\tikzstyle{texte}=[]

\tikzstyle{etiquette}=[midway]

\def\DistanceInterNiveaux{5}
\def\DistanceInterFeuilles{5}

\def\NiveauA{(-0)*\DistanceInterNiveaux}
\def\NiveauB{(-0.7)*\DistanceInterNiveaux}
\def\NiveauC{(-1.4)*\DistanceInterNiveaux}
\def\NiveauD{(-2.1)*\DistanceInterNiveaux}

\def\InterFeuilles{(0.5)*\DistanceInterFeuilles}

\node[noeudn] (A) at ({(0)*\InterFeuilles},{\NiveauA}) {};
\node[texte] (TA) at ({(0)*\InterFeuilles+(-1.5)},{\NiveauA}) {\tiny$v$};

\node[noeudr] (B1) at ({(-2)*\InterFeuilles},{\NiveauB}) {};
\node[texte] (TB1) at ({(-2)*\InterFeuilles+(-1.5)},{\NiveauC}) {\tiny$w$};

\node[noeudn] (B2) at ({(0)*\InterFeuilles},{\NiveauB}) {};
\node[noeudn] (B3) at ({(2)*\InterFeuilles},{\NiveauB}) {};

\node[noeudn] (NEW) at ({(-2)*\InterFeuilles},{\NiveauC}) {};

\node[noeudn] (C11) at ({(-3)*\InterFeuilles},{\NiveauD}) {};
\node[noeudn] (C12) at ({(-2)*\InterFeuilles},{\NiveauD}) {};
\node[noeudn] (C13) at ({(-1)*\InterFeuilles},{\NiveauD}) {};

\draw[fleche] (A)--(B1) node[etiquette] {};
\draw[fleche] (A)--(B2) node[etiquette] {};
\draw[fleche] (A)--(B3) node[etiquette] {};

\draw[fleche] (B1)--(NEW) node[etiquette] {};

\draw[fleche] (NEW)--(C11) node[etiquette] {};
\draw[fleche] (NEW)--(C12) node[etiquette] {};
\draw[fleche] (NEW)--(C13) node[etiquette] {};

\end{tikzpicture}
}

\newcommand{\operationsIB}{
\begin{tikzpicture}[xscale=0.15,yscale=0.15]
\tikzstyle{fleche}=[-,>=latex]
\tikzstyle{noeud}=[fill=white,circle,draw]
\tikzstyle{noeudn}=[fill=black,circle,draw,scale=0.3]
\tikzstyle{noeudb}=[fill=white,circle,scale=0.3]
\tikzstyle{texte}=[]
\tikzstyle{noeudtr}=[fill=white,circle]

\tikzstyle{etiquette}=[midway]

\def\DistanceInterNiveaux{5}
\def\DistanceInterFeuilles{5}

\def\NiveauA{(-0)*\DistanceInterNiveaux}
\def\NiveauB{(-0.7)*\DistanceInterNiveaux}
\def\NiveauC{(-1.4)*\DistanceInterNiveaux}
\def\NiveauD{(-2.1)*\DistanceInterNiveaux}
\def\NiveauE{(-2.8)*\DistanceInterNiveaux}
\def\NiveauF{(-3.5)*\DistanceInterNiveaux}
\def\NiveauG{(-4.2)*\DistanceInterNiveaux}

\def\InterFeuilles{(0.5)*\DistanceInterFeuilles}

\node[noeudn] (A) at ({(0)*\InterFeuilles},{\NiveauA}) {};

\node[noeudn] (B1) at ({(-2)*\InterFeuilles},{\NiveauB}) {};
\node[noeudn] (B2) at ({(0)*\InterFeuilles},{\NiveauB}) {};
\node[noeudn] (B3) at ({(2)*\InterFeuilles},{\NiveauB}) {};

\draw[fleche] (A)--(B1) node[etiquette] {};
\draw[fleche] (A)--(B2) node[etiquette] {};
\draw[fleche] (A)--(B3) node[etiquette] {};

\draw (B1)--({(-3)*\InterFeuilles},{\NiveauD}) node[etiquette] {};
\draw (B1)--({(-1)*\InterFeuilles},{\NiveauD}) node[etiquette] {};
\draw ({(-3)*\InterFeuilles},{\NiveauD})--({(-1)*\InterFeuilles},{\NiveauD}) node[etiquette] {};

\draw (B2)--({(-1)*\InterFeuilles},{\NiveauF}) node[etiquette] {};
\draw (B2)--({(1)*\InterFeuilles},{\NiveauF}) node[etiquette] {};
\draw ({(-1)*\InterFeuilles},{\NiveauF})--({(1)*\InterFeuilles},{\NiveauF}) node[etiquette] {};

\node[texte] (Tcheck) at (0,-20) {\checkmark};
\draw[red,thick] \boundellipse{-2.5,-1.7}{0.5}{0.5};

\node[noeudtr] (AAA) at (-3,-21) {};
\end{tikzpicture}
}

\newcommand{\operationsIC}{
\begin{tikzpicture}[xscale=0.15,yscale=0.15]
\tikzstyle{fleche}=[-,>=latex]
\tikzstyle{noeud}=[fill=white,circle,draw]
\tikzstyle{noeudn}=[fill=black,circle,draw,scale=0.3]
\tikzstyle{noeudb}=[fill=white,circle,scale=0.3]
\tikzstyle{texte}=[]
\tikzstyle{noeudtr}=[fill=white,circle]

\tikzstyle{etiquette}=[midway]

\def\DistanceInterNiveaux{5}
\def\DistanceInterFeuilles{5}

\def\NiveauA{(-0)*\DistanceInterNiveaux}
\def\NiveauB{(-0.7)*\DistanceInterNiveaux}
\def\NiveauC{(-1.4)*\DistanceInterNiveaux}
\def\NiveauD{(-2.1)*\DistanceInterNiveaux}
\def\NiveauE{(-2.8)*\DistanceInterNiveaux}
\def\NiveauF{(-3.5)*\DistanceInterNiveaux}
\def\NiveauG{(-4.2)*\DistanceInterNiveaux}

\def\InterFeuilles{(0.5)*\DistanceInterFeuilles}

\node[noeudn] (A) at ({(0)*\InterFeuilles},{\NiveauA}) {};

\node[noeudn] (B1) at ({(-2)*\InterFeuilles},{\NiveauB}) {};
\node[noeudn] (B2) at ({(0)*\InterFeuilles},{\NiveauB}) {};
\node[noeudn] (B3) at ({(2)*\InterFeuilles},{\NiveauB}) {};

\draw[fleche] (A)--(B1) node[etiquette] {};
\draw[fleche] (A)--(B2) node[etiquette] {};
\draw[fleche] (A)--(B3) node[etiquette] {};

\draw (B1)--({(-3)*\InterFeuilles},{\NiveauD}) node[etiquette] {};
\draw (B1)--({(-1)*\InterFeuilles},{\NiveauD}) node[etiquette] {};
\draw ({(-3)*\InterFeuilles},{\NiveauD})--({(-1)*\InterFeuilles},{\NiveauD}) node[etiquette] {};

\draw (B2)--({(-1)*\InterFeuilles},{\NiveauF}) node[etiquette] {};
\draw (B2)--({(1)*\InterFeuilles},{\NiveauF}) node[etiquette] {};
\draw ({(-1)*\InterFeuilles},{\NiveauF})--({(1)*\InterFeuilles},{\NiveauF}) node[etiquette] {};

\node[texte] (Tcheck) at (0,-20) {\xmark};
\draw[red,thick] \boundellipse{0,-1.7}{0.5}{0.5};
\node[noeudtr] (AAA) at (-3,-21) {};

\end{tikzpicture}
}

\newcommand{\operationsID}{
\begin{tikzpicture}[xscale=0.15,yscale=0.15]
\tikzstyle{fleche}=[-,>=latex]

\tikzstyle{noeudtr}=[fill=white,circle]
\tikzstyle{noeud}=[fill=white,circle,draw]
\tikzstyle{noeudn}=[fill=black,circle,draw,scale=0.3]
\tikzstyle{noeudb}=[fill=white,draw=red,circle,scale=0.3,thick]
\tikzstyle{texte}=[]

\tikzstyle{etiquette}=[midway]

\def\DistanceInterNiveaux{5}
\def\DistanceInterFeuilles{5}

\def\NiveauA{(-0)*\DistanceInterNiveaux}
\def\NiveauB{(-0.7)*\DistanceInterNiveaux}
\def\NiveauC{(-1.4)*\DistanceInterNiveaux}
\def\NiveauD{(-2.1)*\DistanceInterNiveaux}
\def\NiveauE{(-2.8)*\DistanceInterNiveaux}
\def\NiveauF{(-3.5)*\DistanceInterNiveaux}
\def\NiveauG{(-4.2)*\DistanceInterNiveaux}

\def\InterFeuilles{(0.5)*\DistanceInterFeuilles}

\node[noeudn] (A) at ({(0)*\InterFeuilles},{\NiveauA}) {};

\node[noeudn] (B1) at ({(-1)*\InterFeuilles},{\NiveauB}) {};
\node[noeudb] (B2) at ({(2.5)*\InterFeuilles},{\NiveauB}) {};
\node[noeudn] (B3) at ({(1)*\InterFeuilles},{\NiveauB}) {};

\draw[fleche] (A)--(B1) node[etiquette] {};
\draw[fleche] (A)--(B3) node[etiquette] {};
\draw[dash pattern=on 1mm off 0.5mm,red] (A)--(B2) node[etiquette] {};

\draw (B1)--({(-2.5)*\InterFeuilles},{\NiveauE}) node[etiquette] {};
\draw (B1)--({(0.5)*\InterFeuilles},{\NiveauE}) node[etiquette] {};
\draw ({(-2.5)*\InterFeuilles},{\NiveauE})--({(0.5)*\InterFeuilles},{\NiveauE}) node[etiquette] {};

\draw[red] (B2)--({(1.5)*\InterFeuilles},{\NiveauD}) node[etiquette] {};
\draw[red] (B2)--({(3.5)*\InterFeuilles},{\NiveauD}) node[etiquette] {};
\draw[red] ({(1.5)*\InterFeuilles},{\NiveauD})--({(3.5)*\InterFeuilles},{\NiveauD}) node[etiquette] {};

\node[texte] (Tcheck) at (0,-20) {\checkmark};
\node[noeudtr] (AAA) at (-3,-21) {};

\end{tikzpicture}
}

\newcommand{\operationsIE}{
\begin{tikzpicture}[xscale=0.15,yscale=0.15]
\tikzstyle{fleche}=[-,>=latex]

\tikzstyle{noeud}=[fill=white,circle,draw]
\tikzstyle{noeudn}=[fill=black,circle,draw,scale=0.3]
\tikzstyle{noeudb}=[fill=white,draw=red,circle,scale=0.3,thick]
\tikzstyle{texte}=[]

\tikzstyle{noeudtr}=[fill=white,circle]

\tikzstyle{etiquette}=[midway]

\def\DistanceInterNiveaux{5}
\def\DistanceInterFeuilles{5}

\def\NiveauA{(-0)*\DistanceInterNiveaux}
\def\NiveauB{(-0.7)*\DistanceInterNiveaux}
\def\NiveauC{(-1.4)*\DistanceInterNiveaux}
\def\NiveauD{(-2.1)*\DistanceInterNiveaux}
\def\NiveauE{(-2.8)*\DistanceInterNiveaux}
\def\NiveauF{(-3.5)*\DistanceInterNiveaux}
\def\NiveauG{(-4.2)*\DistanceInterNiveaux}

\def\InterFeuilles{(0.5)*\DistanceInterFeuilles}

\node[noeudn] (A) at ({(0)*\InterFeuilles},{\NiveauA}) {};

\node[noeudn] (B1) at ({(-1)*\InterFeuilles},{\NiveauB}) {};
\node[noeudb] (B2) at ({(2.5)*\InterFeuilles},{\NiveauB}) {};
\node[noeudn] (B3) at ({(1)*\InterFeuilles},{\NiveauB}) {};

\draw[fleche] (A)--(B1) node[etiquette] {};
\draw[fleche] (A)--(B3) node[etiquette] {};
\draw[dash pattern=on 1mm off 0.5mm,red] (A)--(B2) node[etiquette] {};

\draw (B1)--({(-2.5)*\InterFeuilles},{\NiveauE}) node[etiquette] {};
\draw (B1)--({(0.5)*\InterFeuilles},{\NiveauE}) node[etiquette] {};
\draw ({(-2.5)*\InterFeuilles},{\NiveauE})--({(0.5)*\InterFeuilles},{\NiveauE}) node[etiquette] {};

\draw[red] (B2)--({(1.5)*\InterFeuilles},{\NiveauF}) node[etiquette] {};
\draw[red] (B2)--({(3.5)*\InterFeuilles},{\NiveauF}) node[etiquette] {};
\draw[red] ({(1.5)*\InterFeuilles},{\NiveauF})--({(3.5)*\InterFeuilles},{\NiveauF}) node[etiquette] {};

\node[texte] (Tcheck) at (0,-20) {\xmark};

\node[noeudtr] (AAA) at (-3,-21) {};

\end{tikzpicture}
}

\newcommand{\operationsDA}{
\begin{tikzpicture}[xscale=0.15,yscale=0.15]
\tikzstyle{fleche}=[-,>=latex]
\tikzstyle{noeud}=[fill=white,circle,draw]
\tikzstyle{noeudn}=[fill=black,circle,draw,scale=0.3]
\tikzstyle{noeudb}=[fill=white,circle,scale=0.3]
\tikzstyle{noeudtr}=[fill=white,circle]
\tikzstyle{texte}=[]

\tikzstyle{etiquette}=[midway]

\def\DistanceInterNiveaux{5}
\def\DistanceInterFeuilles{5}

\def\NiveauA{(-0)*\DistanceInterNiveaux}
\def\NiveauB{(-0.7)*\DistanceInterNiveaux}
\def\NiveauC{(-1.4)*\DistanceInterNiveaux}
\def\NiveauD{(-2.1)*\DistanceInterNiveaux}
\def\NiveauE{(-2.8)*\DistanceInterNiveaux}
\def\NiveauF{(-3.5)*\DistanceInterNiveaux}
\def\NiveauG{(-4.2)*\DistanceInterNiveaux}

\def\InterFeuilles{(0.5)*\DistanceInterFeuilles}

\node[noeudn] (A) at ({(0)*\InterFeuilles},{\NiveauA}) {};

\node[noeudn] (B1) at ({(-2)*\InterFeuilles},{\NiveauB}) {};
\node[texte] (BB1) at ({(-2)*\InterFeuilles},{\NiveauB}) {${\color{red}{\Cross}}$};

\node[noeudn] (B3) at ({(2)*\InterFeuilles},{\NiveauB}) {};

\node[noeudn] (C1) at ({(-2)*\InterFeuilles},{\NiveauC}) {};
\node[noeudn] (C3) at ({(2)*\InterFeuilles},{\NiveauC}) {};

\draw[fleche] (A)--(B1) node[etiquette] {};
\draw[fleche] (A)--(B3) node[etiquette] {};
\draw[fleche] (B1)--(C1) node[etiquette] {};
\draw[fleche] (B3)--(C3) node[etiquette] {};

\draw (C1)--({(-2.5)*\InterFeuilles},{\NiveauE}) node[etiquette] {};
\draw (C1)--({(-1.5)*\InterFeuilles},{\NiveauE}) node[etiquette] {};
\draw ({(-2.5)*\InterFeuilles},{\NiveauE})--({(-1.5)*\InterFeuilles},{\NiveauE}) node[etiquette] {};

\draw (C3)--({(0.5)*\InterFeuilles},{\NiveauF}) node[etiquette] {};
\draw (C3)--({(3.5)*\InterFeuilles},{\NiveauF}) node[etiquette] {};
\draw ({(0.5)*\InterFeuilles},{\NiveauF})--({(3.5)*\InterFeuilles},{\NiveauF}) node[etiquette] {};

\node[noeudtr] (AAA) at (-3,-21) {};
\node[texte] (Tcheck) at (0,-20) {\checkmark};

\end{tikzpicture}
}

\newcommand{\operationsDB}{
\begin{tikzpicture}[xscale=0.15,yscale=0.15]
\tikzstyle{fleche}=[-,>=latex]
\tikzstyle{noeud}=[fill=white,circle,draw]
\tikzstyle{noeudn}=[fill=black,circle,draw,scale=0.3]
\tikzstyle{noeudb}=[fill=white,circle,scale=0.3]
\tikzstyle{noeudtr}=[fill=white,circle]
\tikzstyle{texte}=[]

\tikzstyle{etiquette}=[midway]

\def\DistanceInterNiveaux{5}
\def\DistanceInterFeuilles{5}

\def\NiveauA{(-0)*\DistanceInterNiveaux}
\def\NiveauB{(-0.7)*\DistanceInterNiveaux}
\def\NiveauC{(-1.4)*\DistanceInterNiveaux}
\def\NiveauD{(-2.1)*\DistanceInterNiveaux}
\def\NiveauE{(-2.8)*\DistanceInterNiveaux}
\def\NiveauF{(-3.5)*\DistanceInterNiveaux}
\def\NiveauG{(-4.2)*\DistanceInterNiveaux}

\def\InterFeuilles{(0.5)*\DistanceInterFeuilles}

\node[noeudn] (A) at ({(0)*\InterFeuilles},{\NiveauA}) {};

\node[noeudn] (B1) at ({(-2)*\InterFeuilles},{\NiveauB}) {};

\node[noeudn] (B3) at ({(2)*\InterFeuilles},{\NiveauB}) {};
\node[texte] (BB3) at ({(2)*\InterFeuilles},{\NiveauB}) {${\color{red}{\Cross}}$};

\node[noeudn] (C1) at ({(-2)*\InterFeuilles},{\NiveauC}) {};
\node[noeudn] (C3) at ({(2)*\InterFeuilles},{\NiveauC}) {};

\draw[fleche] (A)--(B1) node[etiquette] {};
\draw[fleche] (A)--(B3) node[etiquette] {};
\draw[fleche] (B1)--(C1) node[etiquette] {};
\draw[fleche] (B3)--(C3) node[etiquette] {};

\draw (C1)--({(-2.5)*\InterFeuilles},{\NiveauE}) node[etiquette] {};
\draw (C1)--({(-1.5)*\InterFeuilles},{\NiveauE}) node[etiquette] {};
\draw ({(-2.5)*\InterFeuilles},{\NiveauE})--({(-1.5)*\InterFeuilles},{\NiveauE}) node[etiquette] {};

\draw (C3)--({(0.5)*\InterFeuilles},{\NiveauF}) node[etiquette] {};
\draw (C3)--({(3.5)*\InterFeuilles},{\NiveauF}) node[etiquette] {};
\draw ({(0.5)*\InterFeuilles},{\NiveauF})--({(3.5)*\InterFeuilles},{\NiveauF}) node[etiquette] {};

\node[noeudtr] (AAA) at (-3,-21) {};
\node[texte] (Tcheck) at (0,-20) {\xmark};

\end{tikzpicture}
}

\newcommand{\operationsDC}{
\begin{tikzpicture}[xscale=0.15,yscale=0.15]
\tikzstyle{fleche}=[-,>=latex]
\tikzstyle{noeud}=[fill=white,circle,draw]
\tikzstyle{noeudn}=[fill=black,circle,draw,scale=0.3]
\tikzstyle{noeudb}=[fill=white,circle,scale=0.3]
\tikzstyle{noeudtr}=[fill=white,circle]
\tikzstyle{texte}=[]

\tikzstyle{etiquette}=[midway]

\def\DistanceInterNiveaux{5}
\def\DistanceInterFeuilles{5}

\def\NiveauA{(-0)*\DistanceInterNiveaux}
\def\NiveauB{(-0.7)*\DistanceInterNiveaux}
\def\NiveauC{(-1.4)*\DistanceInterNiveaux}
\def\NiveauD{(-2.1)*\DistanceInterNiveaux}
\def\NiveauE{(-2.8)*\DistanceInterNiveaux}
\def\NiveauF{(-3.5)*\DistanceInterNiveaux}
\def\NiveauG{(-4.2)*\DistanceInterNiveaux}

\def\InterFeuilles{(0.5)*\DistanceInterFeuilles}

\node[noeudn] (A) at ({(0)*\InterFeuilles},{\NiveauA}) {};

\node[noeudn] (B1) at ({(-2)*\InterFeuilles},{\NiveauB}) {};

\node[noeudn] (B3) at ({(2)*\InterFeuilles},{\NiveauB}) {};

\draw[fleche] (A)--(B1) node[etiquette] {};
\draw[fleche] (A)--(B3) node[etiquette] {};

\draw (B1)--({(-3)*\InterFeuilles},{\NiveauE}) node[etiquette] {};
\draw (B1)--({(-1)*\InterFeuilles},{\NiveauE}) node[etiquette] {};
\draw ({(-3)*\InterFeuilles},{\NiveauE})--({(-1)*\InterFeuilles},{\NiveauE}) node[etiquette] {};

\draw (B3)--({(1)*\InterFeuilles},{\NiveauF}) node[etiquette] {};
\draw (B3)--({(3)*\InterFeuilles},{\NiveauF}) node[etiquette] {};
\draw ({(1)*\InterFeuilles},{\NiveauF})--({(3)*\InterFeuilles},{\NiveauF}) node[etiquette] {};

\node[texte] (BB3) at ({(-2)*\InterFeuilles},{\NiveauC-2}) {${\color{red}{\BigCross}}$};

\node[noeudtr] (AAA) at (-3,-21) {};
\node[texte] (Tcheck) at (0,-20) {\checkmark};

\end{tikzpicture}
}

\newcommand{\operationsDD}{
\begin{tikzpicture}[xscale=0.15,yscale=0.15]
\tikzstyle{fleche}=[-,>=latex]
\tikzstyle{noeud}=[fill=white,circle,draw]
\tikzstyle{noeudn}=[fill=black,circle,draw,scale=0.3]
\tikzstyle{noeudb}=[fill=white,circle,scale=0.3]
\tikzstyle{noeudtr}=[fill=white,circle]
\tikzstyle{texte}=[]

\tikzstyle{etiquette}=[midway]

\def\DistanceInterNiveaux{5}
\def\DistanceInterFeuilles{5}

\def\NiveauA{(-0)*\DistanceInterNiveaux}
\def\NiveauB{(-0.7)*\DistanceInterNiveaux}
\def\NiveauC{(-1.4)*\DistanceInterNiveaux}
\def\NiveauD{(-2.1)*\DistanceInterNiveaux}
\def\NiveauE{(-2.8)*\DistanceInterNiveaux}
\def\NiveauF{(-3.5)*\DistanceInterNiveaux}
\def\NiveauG{(-4.2)*\DistanceInterNiveaux}

\def\InterFeuilles{(0.5)*\DistanceInterFeuilles}

\node[noeudn] (A) at ({(0)*\InterFeuilles},{\NiveauA}) {};

\node[noeudn] (B1) at ({(-2)*\InterFeuilles},{\NiveauB}) {};

\node[noeudn] (B3) at ({(2)*\InterFeuilles},{\NiveauB}) {};

\draw[fleche] (A)--(B1) node[etiquette] {};
\draw[fleche] (A)--(B3) node[etiquette] {};

\draw (B1)--({(-3)*\InterFeuilles},{\NiveauE}) node[etiquette] {};
\draw (B1)--({(-1)*\InterFeuilles},{\NiveauE}) node[etiquette] {};
\draw ({(-3)*\InterFeuilles},{\NiveauE})--({(-1)*\InterFeuilles},{\NiveauE}) node[etiquette] {};

\draw (B3)--({(1)*\InterFeuilles},{\NiveauF}) node[etiquette] {};
\draw (B3)--({(3)*\InterFeuilles},{\NiveauF}) node[etiquette] {};
\draw ({(1)*\InterFeuilles},{\NiveauF})--({(3)*\InterFeuilles},{\NiveauF}) node[etiquette] {};

\node[texte] (BB3) at ({(2)*\InterFeuilles},{\NiveauD}) {${\color{red}{\BigCross}}$};

\node[noeudtr] (AAA) at (-3,-21) {};
\node[texte] (Tcheck) at (0,-20) {\xmark};

\end{tikzpicture}
}

\title{\Large\textsc{Nearest Embedded and Embedding Self-Nested Trees}}

\author{Romain Aza\"is}

\date{\small Laboratoire Reproduction et D\'eveloppement des Plantes, Univ Lyon,\\
ENS de Lyon, UCB Lyon 1, CNRS, INRA, Inria, F-69342, Lyon, France.}

\begin{document}

\maketitle

\vspace{0.25cm}


{\small
\begin{center}\textbf{Abstract}\end{center}

\vspace{-0.1cm}

\noindent
Self-nested trees present a systematic form of redundancy in their subtrees and thus achieve optimal compression rates by DAG compression. A method for quantifying the degree of self-similarity of plants through self-nested trees has been introduced by Godin and Ferraro in 2010. The procedure consists in computing a self-nested approximation, called the nearest embedding self-nested tree, that both embeds the plant and is the closest to it. In this paper, we propose a new algorithm that computes the nearest embedding self-nested tree with a smaller overall complexity, but also the nearest embedded self-nested tree. We show from simulations that the latter is mostly the closest to the initial data, which suggests that this better approximation should be used as a privileged measure of the degree of self-similarity of plants.

\smallskip

\noindent
\textbf{keywords:} unordered trees; self-nested trees; approximation of trees; structural self-similarity
}


\vspace{0.25cm}

\section{Introduction}

\begin{figure}
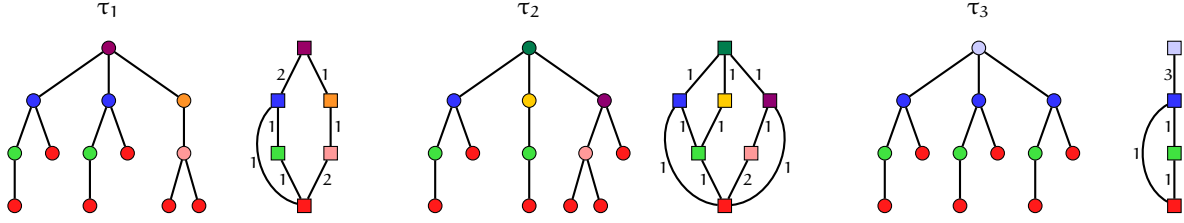

\centering
\treeTA\qquad~\treeTC\qquad~\treeTB
\caption{Trees and their DAG reduction. In the tree, roots of isomorphic
subtrees are colored identically. In the DAG, vertices are equivalence classes colored according
to the class of isomorphic subtrees that they represent.}
\label{fig:1}
\end{figure}

Trees form a wide family of combinatorial objects that offers many application fields, e.g., plant modeling and XML files analysis. Modern databases are huge and thus stored in compressed form. Compression methods take advantage of repeated substructures appearing in the tree. As explained in \cite{Bille2015166}, one often considers the following two types of repeated substructures: subtree repeat (used in DAG compression \cite{BM,Buneman:2003:PQC:1315451.1315465,frick2003query,GF2010}) and tree pattern repeat (exploited in tree grammars \cite{Busatto:2008:EMR:1370308.1370503,Lohrey:2006:CTA:1217607.1217615} and top tree compression \cite{Bille2015166}). We restrict ourselves to DAG compression of unordered rooted trees, which consists in building a \underline{D}irected \underline{A}cyclic \underline{G}raph (DAG) that represents a tree without the redundancy of its identical subtrees (see {Fig.\,\ref{fig:1}}). Two different algorithms exist for computing the DAG reduction of a tree $\tau$ \cite[2.2 Computing Tree Reduction]{GF2010}, which share the same time-complexity in $O(\#\mathcal{V}(\tau)^2\times\mathcal{D}(\tau)\times\log(\mathcal{D}(\tau)))$ where $\mathcal{V}(\tau)$ denotes the set of vertices of $\tau$ and $\mathcal{D}(\tau)$ its outdegree.

\medskip

\noindent
Trees that are the most compressed by DAG compression present the highest level of redundancy in their subtrees: all the subtrees of a given height are isomorphic. In this case, the DAG related to a tree $\tau$ is linear, i.e., there exists a path going through all vertices, with exactly $\mathcal{H}(\tau)+1$ vertices, $\mathcal{H}(\tau)$ denoting the height of $\tau$, which is the minimal number of vertices among trees of this height (see $\tau_3$ in {Fig.\,\ref{fig:1}}). This family of trees has been introduced in \cite{greenlaw96} as the first interesting class of trees for which the subtree isomorphism problem is in NC$^2$. It has been known under the name of nested trees \cite{greenlaw96} and next self-nested trees \cite{GF2010} to insist on their recursive structure and their proximity to the notion of structural self-similarity.

\medskip

\begin{figure}[!h]
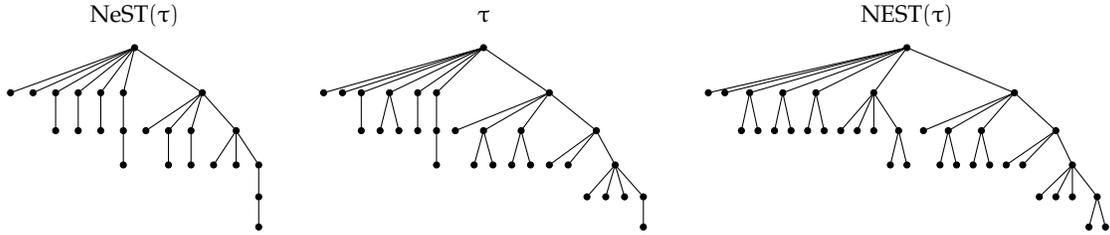

\centering
\treeintro
\caption{A tree $\tau$ (middle) with $30$ nodes and its approximations $\treeNeST(\tau)$ (left) with $24$ nodes and $\treeNEST(\tau)$ (right) with $37$ nodes.}
\label{fig:2}
\end{figure}

\noindent
The authors of \cite{GF2010} are interested in capturing the self-similarity of plants through self-nested trees. They propose to construct a self-nested tree that minimizes the distance of the original tree to the set of self-nested trees that embed the initial tree. The distance to this \underline{N}earest \underline{E}mbedding \underline{S}elf-nested \underline{T}ree (NEST) is then used to quantify the self-nestedness of the tree and thus its structural self-similarity (see $\tau$ and $\treeNEST(\tau)$ in Fig.\,\ref{fig:2}). The main result of \cite[Theorem 1 and E.\,NEST Algorithm]{GF2010} is an algorithm that computes the NEST of a tree $\tau$ from its DAG reduction in $O(\mathcal{H}(\tau)^2\times\mathcal{D}(\tau))$.

\noindent
The goal of the present article is threefold. We aim at proposing a new and more explicit algorithm that computes the NEST of a tree $\tau$ with the same time-complexity $O(\mathcal{H}(\tau)^2\times\mathcal{D}(\tau))$ as in \cite{GF2010} but that takes as input the height profile of $\tau$ and not its DAG reduction. We establish that the height profile of a tree $\tau$ can be computed in $O(\#\mathcal{V}(\tau)\times\mathcal{D}(\tau))$ reducing the overall complexity of a linear factor. Based on this work, we also provide an algorithm in $O(\mathcal{H}(\tau)^2)$ that computes the \underline{N}earest \underline{e}mbedded \underline{S}elf-nested \underline{T}ree (NeST) of a tree $\tau$ (see $\tau$ and $\treeNeST(\tau)$ in Fig.\,\ref{fig:2}). Finally, we show from numerical simulations that the distance of a tree $\tau$ to its NeST is much lower than the distance to its NEST. The NeST is most of the time a better approximation of a tree than the NEST and thus should be privileged to quantify the degree of self-nestedness of plants.

\medskip

\noindent
The paper is organized as follows. The structures of interest in this paper, namely unordered trees, DAG compression and self-nested trees, are defined in Section \ref{s:2}. Section \ref{s:3} is dedicated to the definition and the study of the height profile of a tree. The approximation algorithms are presented in Section \ref{s:4}. We give a new insight on the definitions of the NEST and of the NeST in Subsection \ref{s:4:ss:def}. Our NEST algorithm is presented in Subsection \ref{s:4:ss:NEST}, while the NeST algorithm is given in Subsection \ref{s:4:ss:NeST}. Section \ref{s:5} is devoted to simulations. We state that the NeST is mostly a better approximation of a tree than the NEST in Subsection \ref{ss:51}. An application to a real rice panicle is presented in Subsection \ref{ss:52}. A summary of the paper and concluding remarks can be found in Section~\ref{s:6}. All the figures and numerical experiments presented in the article have been made with the \verb+Python+ library \verb+treex+ \cite{Azais2019treex}.


\section{Preliminaries}
\label{s:2}

\subsection{Unordered rooted trees}

A rooted tree $\tau$ is a connected graph containing no cycle, that is, without chain from any vertex $v$ to itself, and such that there exists a unique vertex $\mathcal{R}(\tau)$, called the root, which has no parent, and any vertex different from the root has exactly one parent. The leaves of $\tau$ are all the vertices without children. The set of vertices of $\tau$ is denoted by $\mathcal{V}(\tau)$. The height of a vertex $v$ may be recursively defined as $\mathcal{H}(v)=0$ if $v$ is a leaf of $\tau$ and
$$\mathcal{H}(v)=1+\max_{w\in\mathcal{C}_\tau(v)}\mathcal{H}(w)$$
otherwise, $\mathcal{C}_\tau(v)$ denoting the set of children of $v$ in $\tau$. The height of the tree $\tau$ is defined as the height of its root, $\mathcal{H}(\tau)=\mathcal{H}(\mathcal{R}(\tau))$.The outdegree $\mathcal{D}(\tau)$ of $\tau$ is the maximal branching factor that can be found in $\tau$, that is
$$\mathcal{D}(\tau)=\max_{v\in\tau}\#\mathcal{C}_\tau(v).$$
A subtree $\tau[v]$ rooted in $v$ is a particular connected subgraph of $\tau$. Precisely, $\tau[v]=(V[v],E[v])$ where $V[v]$ is the set of the descendants of $v$ in $\tau$ and $E[v]$ is defined as
$$E[v]=\left\{(\xi,\xi')\in\mathcal{E}(\tau)~:~\xi\in V[v],\,\xi'\in V[v]\right\},$$
with $\mathcal{E}(\tau)$ the set of edges of $\tau$.

\medskip

\noindent
In all the sequel, we consider unordered rooted trees for which the order among the sibling vertices of any vertex is not significant. A precise characterization is obtained from the additional definition of isomorphic trees. Let $\tau$ and $\theta$ two rooted trees. A one-to-one correspondence $\varphi:\mathcal{V}(\tau)\to\mathcal{V}(\theta)$ is called a tree isomorphism if, for any edge $(v,w)\in\mathcal{E}(\tau)$, $(\varphi(v),\varphi(w))\in\mathcal{E}(\theta)$. Structures $\tau_1$ and $\tau_2$ are called isomorphic trees whenever there exists a tree isomorphism between them. One can determine if two $n$-vertex trees are isomorphic in $O(n)$ \cite[Example 3.2 and Theorem 3.3]{Aho:1974:DAC:578775}. The existence of a tree isomorphism defines an equivalence relation on the set of rooted trees. The class of unordered rooted trees is the set of equivalence classes for this relation, i.e., the quotient set of rooted trees by the existence of a tree isomorphism.

\subsection{DAG compression}

Now we consider the equivalence relation ``existence of a tree isomorphism'' on the set of the subtrees of a tree $\tau$. We consider the quotient graph $\mathcal{Q}(\tau)=(V,E)$ obtained from $\tau$ using this equivalence relation. $V$ is the set of equivalence classes on the subtrees of $\tau$, while $E$ is a set of pairs of equivalence classes $(C_1,C_2)$ such that $\mathcal{R}(C_2)\in\mathcal{C}_\tau(\mathcal{R}(C_1))$ up to an isomorphism. The graph $\mathcal{Q}(\tau)$ is a DAG \cite[Proposition 1]{GF2010} that is a connected directed graph without path from any vertex $v$ to itself.

\medskip

\noindent
Let $(C_1,C_2)$ be an edge of the DAG $\mathcal{Q}(\tau)$. We define $N(C_1,C_2)$ as the number of occurrences of a tree of $C_2$ just below the root of any tree of $C_1$. The tree reduction $\mathcal{R}(\tau)$ is defined as the quotient graph $\mathcal{Q}(\tau)$ augmented with labels $N(C_1,C_2)$ on its edges \cite[Definition 3 (Reduction of a tree)]{GF2010}. Intuitively, the graph $\mathcal{R}(\tau)$ represents the original tree $\tau$ without its structural redundancies (see Fig.\,\ref{fig:1}).

\subsection{Self-nested trees}

A tree $\tau$ is called self-nested \cite[III. Self-nested trees]{GF2010} if for any pair of vertices $v$ and $w$, either the subtrees $\tau[v]$ and $\tau[w]$ are isomorphic, or one is (isomorphic to) a subtree of the other. This characterization of self-nested trees is equivalent to the following statement: for any pair of vertices $v$ and $w$ such that $\mathcal{H}(v)=\mathcal{H}(w)$, $\tau[x]=\tau[y]$, i.e., all the subtrees of the same height are isomorphic.

\medskip

\noindent
Linear DAGs are DAGs containing at least one path that goes through all their vertices. They are closely connected with self-nested trees by virtue of the following result.

\begin{proposition}[Godin and Ferraro \cite{GF2010}]
\label{prop:equivalence}
A tree $\tau$ is self-nested if and only if its reduction $\mathcal{R}(\tau)$ is a linear~DAG.
\end{proposition}

\noindent
This result proves that self-nested trees achieve optimal compression rates among trees of the same height whatever their number of nodes (compare $\tau_3$ with $\tau_1$ and $\tau_2$ in Fig.\,\ref{fig:1}). Indeed, $\mathcal{R}(\tau)$ has at least $\mathcal{H}(\tau)+1$ nodes and the inequality is saturated if and only if $\tau$ is self-nested.


\section{Height profile of the tree structure}
\label{s:3}

\subsection{Definition and complexity}

This section is devoted to the definition of the height profile $\rho_\tau$ of a tree $\tau$ and to the presentation of an algorithm to calculate it. In the sequel, we assume that the tree $\tau$ is always traversed in the same order, depth-first search to set the ideas down. In particular, when vectors are indexed by nodes of $\tau$ sharing the same property, the order of the vector is important and should be always the same.

\medskip

\noindent
Given a vertex $v\in\mathcal{V}(\tau)$,
$$\gamma_{h}(v) = \#\,\{v'\in\mathcal{C}_\tau(v)~\!:~\!\mathcal{H}(\tau[v'])=h\}$$
is the number of subtrees of height $h$ directly under $v$. Now, we consider the vector
$$\rho_\tau(h_1,h_2)=\left(\gamma_{h_2}(v)~\!:~\!v\in\mathcal{V}(\tau),\,\mathcal{H}(\tau[v])=h_1\right)$$
made of the concatenation of the integers $\gamma_{h_2}(v)$ over subtrees $\tau[v]$ of height $h_1$ ordered in depth-first search. Consequently, $\rho_\tau$ is an array made of vectors with varying lengths.

\medskip

\noindent
Let $A_1$ and $A_2$ be two arrays for which each entry is a vector. We say that $A_1$ and $A_2$ are equivalent if, for any line $i$, there exists a permutation $\sigma_{i}$ such that, for any column $j$,
$$A_1(i,j) = \sigma_{i}(A_2(i,j)).$$
In particular, $i$ being fixed, all the vectors $A_1(i,j)$ and $A_2(i,j)$ must have the same length. This condition defines an equivalence relation. The height profile of $\tau$ is the array $\rho_\tau$ as an element of the quotient space of arrays of vectors under this equivalence relation. In other words, the vectors $\rho_\tau(h_1,h_2)$, $0\leq h_2<h_1$ and $h_1$ fixed, must be ordered in the same way but the choice of the order is not significant. Finally, it should be already remarked that $\rho_\tau(h_1,h_2) = \emptyset$ when $h_2\geq h_1$ or $h_1>\mathcal{H}(\tau)$. Consequently, the height profile can be reduced to the triangular array
$$ \rho_\tau = \big[\rho_\tau(h_1,h_2)\big]_{0\leq h_2<h_1\leq\mathcal{H}(\tau)} .$$

\medskip

\noindent
The application $\rho_\tau$ provides the distribution of subtrees of height $h_2$ just below the root of subtrees of height $h_1$ for all couples $(h_1,h_2)$, which typically represents the height profile of $\tau$. For clarity's sake, we give the values of $\rho_{\tau_k}$ for the trees $\tau_k$ of {Fig.\,\ref{fig:1}}, coefficient $(i,j)$ of the matrix being $\rho_{\tau_k}(i,j-1)$,
\begin{equation}
\label{eq:profile:ex}
\rho_{\tau_1}=\rho_{\tau_2}=
{\scriptscriptstyle
\left[
\begin{array}{ccc}
\scriptstyle (1,1,2) & \scriptstyle \emptyset &\scriptstyle  \emptyset  \\
\scriptstyle  (0,1,1) & \scriptstyle (1,1,1) & \scriptstyle  \emptyset\\
\scriptstyle (0) &\scriptstyle  (0) &\scriptstyle  (3)
\end{array}
\right]
}
\quad\text{and}\quad\rho_{\tau_3}=
{\scriptscriptstyle
\left[
\begin{array}{ccc}
\scriptstyle (1,1,1) &\scriptstyle \emptyset &\scriptstyle \emptyset  \\
\scriptstyle (1,1,1) & \scriptstyle (1,1,1) & \scriptstyle \emptyset\\
\scriptstyle (0) & \scriptstyle (0) & \scriptstyle (3)
\end{array}
\right]
}.
\end{equation}

\noindent
It should be noticed that the height profile does not contain all the topology of the tree since trees $\tau_1$ and $\tau_2$ of Fig.\,\ref{fig:1} are different but share the same height profile \eqref{eq:profile:ex}. However, the height of a tree $\tau$ can be recovered from its height profile through the relation $\mathcal{H}(\tau)=\dim(\rho_\tau)$, the dimension of $\rho_\tau$ being defined by
$$\dim(\rho_\tau)= \min~\big\{n\geq0~:~\forall\,i\geq0,~\rho_\tau(n+1,i)=\emptyset\big\}.$$

\begin{proposition}\label{prop:hp:complexity}
$\rho_\tau$ can be computed in $O(\#\mathcal{V}(\tau)\times\mathcal{D}(\tau))$-time.
\end{proposition}

\noindent
\textit{Proof.} First, attribute to each node $v\in\mathcal{V}(\tau)$ the height of the subtree $\tau[v]$ with complexity $O(\#\mathcal{V}(\tau))$. Next, traverse the tree in depth-first search in $O(\#\mathcal{V}(\tau))$ and calculate for each vertex $v$ the vector $(\gamma_h(v))_{0\leq h<\mathcal{H}(\tau[v])}$ in $\#\mathcal{C}_\tau(v)\leq\mathcal{D}(\tau)$ operations. Finally, append this vector to $\rho_\tau(\mathcal{H}(\tau[v]),\cdot)$ component by component.\hfill{\color{OrangeRed}{\leafNE}}

\subsection{Relation with self-nested trees}

Self-nested trees are characterized by their height profile in light of the following result.

\begin{proposition}
\label{prop:property:selfnested}
$\tau$ is self-nested if and only if, for any $0\leq h_2<h_1\leq\mathcal{H}(\tau)$, all the components of the vector $\rho_\tau(h_1,h_2)$ are the same (for instance see the profile \eqref{eq:profile:ex} of the tree $\tau_3$ presented in {Fig.\,\ref{fig:1}}). In addition, a self-nested tree $\tau$ can be reconstructed from $\rho_\tau$ (see Algorithm \ref{algo:SN}).
\end{proposition}

\medskip

\noindent
\textit{Proof.} If $\tau$ is self-nested, the $N_{h_1}$ subtrees of height $h_1$ appearing in $\tau$ are isomorphic and thus have the same number $n_{h_1,h_2}$ of subtrees of height $h_2$ just below their root. As a consequence,
$$\rho_\tau(h_1,h_2) = (\underset{N_{h_1}}{\underleftrightarrow{n_{h_1,h_2},\dots,n_{h_1,h_2}}}).$$
The reciprocal result may be established in light of the following lemma which proof presents no difficulty.

\begin{lemma}
If all the subtrees of height $0\leq h<H$ appearing in a tree $\tau$ are isomorphic, and if all the subtrees of height $H$ have the same number of subtrees of height $0\leq h<H$ just below their root, then all the subtrees of height $H$ appearing in $\tau$ are isomorphic.
\end{lemma}

\noindent
All the subtrees of height $1$ in $\tau$ are isomorphic because all the components of $\rho_\tau(1,0)$ are the same. The expected result is shown by induction on the height thanks to the previous lemma which assumptions are satisfied since $\rho_\tau$ always contains vectors for which all the entries are equal.
The previous reasoning also provides a way (presented in Algorithm \ref{algo:SN}) to build a unique (self-nested) tree $\mathcal{T}$ from the height profile $\rho_\tau$. In addition, this is easy to see that $\tau$ and $\mathcal{T}$ are isomorphic.\hfill{\color{OrangeRed}{\leafNE}}

\medskip

\noindent
In order to present the algorithm of reconstruction of a self-nested tree from its height profile, we need to define the restriction of a height profile to some height. Let $p$ be a height profile. The restriction $p_{|_h}$ of $p$ to height $h\geq0$ is the array defined by
$$
\left\{
\begin{array}{lll}
\forall\,1\leq h_1\leq h,&\forall\,h_2\geq0,&p_{|_h}(h_1,h_2) = p(h_1,h_2),\\
\forall\,h_1>h,&\forall\,h_2\geq0,&p_{|_h}(h_1,h_2) = \emptyset .
\end{array}
\right.
$$
Consequently, $\dim(p_{|_h})=\min(dim(p),h)$. A peculiar case is $p_{|_0}$ for which each entry is the empty set and thus $\dim(p_{|_0})=0$. It should be also remarked that there may exist no tree $\tau$ such that $p_{|_h}$ is the height profile of $\tau$.

\smallskip

\SetKwFunction{SN}{SN}%
\SetKwProg{Fn}{Function}{\string:}{}
\begin{algorithm}
\DontPrintSemicolon
\Fn(){\SN{$p$}}{
\KwData{a height profile $p$ such that all the components of $p(h_1,h_2)$ are the same}
\KwResult{the unique self-nested tree $\tau$ such that $\rho_\tau=p$}
$\tau=\bullet$\\
\For{$i$ {\bf{from}} $0$ {\bf{to}} $\dim(p)-1$}{
	add \SN($p_{|_i}$) as child of $\mathcal{R}(\tau)$ $p(\dim(p),i)_1$ times
}
\Return{$\tau$}
}
\caption{Construction of a self-nested tree from its height profile}
\label{algo:SN}
\end{algorithm}

\smallskip

\noindent
As we can see in the proof of Proposition \ref{prop:property:selfnested} or in Algorithm \ref{algo:SN}, the lengths of the vectors $\rho_\tau(h_1,h_2)$ are not significant to reconstruct a self-nested tree $\tau$. Consequently, since all the components of $\rho_\tau(h_1,h_2)$ are the same, we can identify the height profile of a self-nested tree with the integer-valued array $[\rho_\tau(h_1,h_2)_1]$.

\begin{proposition}
The number of nodes of a self-nested tree $\tau$ can be computed from $\rho_\tau$ in $O(\mathcal{H}(\tau)^2)$.
\end{proposition}

\noindent
\textit{Proof.}
By induction on the height, one has $\#\mathcal{V}(\tau)=\mathcal{N}(\mathcal{H}(\tau))$,
where the sequence $\mathcal{N}$ is defined by $\mathcal{N}(0)=1$ (number of nodes of a tree reduced to a root) and,
\begin{equation}
\label{eq:nnodes:sn:hp}
\forall\,1\leq H\leq\mathcal{H}(\tau),~\mathcal{N}(H) = 1+\sum_{h=0}^{H-1} \rho_\tau(H,h)\mathcal{N}(h).
\end{equation}
The number of operations required to compute $\mathcal{N}(\mathcal{H}(\tau))$ is of order $O(\mathcal{H}(\tau)^2)$.\hfill{\color{OrangeRed}{\leafNE}}

\bigskip

\noindent
The authors of \cite[Proposition 6]{GF2010} calculate the number of nodes of a tree (self-nested or not) from its DAG reduction by a formula very similar to \eqref{eq:nnodes:sn:hp}, and which achieve the same complexity on self-nested trees. As mentioned before, a tree can not be recovered from its height profile in general, thus we can not expect such a result from the height profile of any tree.


\section{Approximation algorithms}
\label{s:4}

\subsection{Definitions}
\label{s:4:ss:def}

\subsubsection{Editing operations}

We shall define the NEST and the NeST of a tree $\tau$. As in \cite[eq.\,(5)]{GF2010}, we ask these approximations to be consistent with Zhang's edit distance between unordered trees \cite{Zhang1996} denoted $D_Z$ in this paper. Thus, as in \cite[2.2 Editing Operations]{Zhang1996}, we consider the following two types of editing operations: adding a node and deleting a node. Deleting a node $w$ means making the children of $w$ become the children of the parent $v$ of $w$ and then removing $w$ (see Fig.\,\ref{fig:delnode}). Adding $w$ as a child of $v$ will make $w$ the parent of a subset of the current children of $v$ (see Fig.\,\ref{fig:insertnode}).

\bigskip
\bigskip

\begin{minipage}{.40\textwidth}
\centering
\treeTdelA\quad$\stackrel{\longrightarrow}{\text{\textcolor{white}{$\displaystyle\sum$}}}$\quad\treeTdelB
\captionsetup{type=figure}

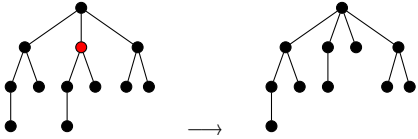
\captionof{figure}{Deleting a node.}
\label{fig:delnode}
\end{minipage}
\quad\qquad
\begin{minipage}{.40\textwidth}\centering
\centering
\treeTinsertleafA\quad$\stackrel{\longrightarrow}{\text{\textcolor{white}{$\displaystyle\sum$}}}$\quad\treeTinsertleafC
\captionsetup{type=figure}

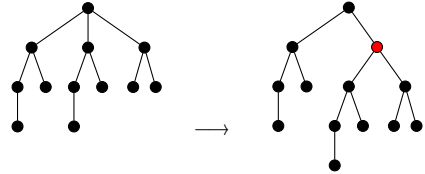
\captionof{figure}{Inserting a node.}
\label{fig:insertnode}
\centering
\end{minipage}

\subsubsection{Constrained editing operations}

Zhang's edit distance is defined from the above editing operations and from constrained mappings between trees \cite[3.1 Constrained Edit Distance Mappings]{Zhang1996}. A constrained mapping between two trees $\tau$ and $\theta$ is a mapping \cite[2.3.2 Editing Distance Mappings]{Zhang1996}, i.e., a one-to-one correspondence $\varphi$ from a subset of $\mathcal{V}(\tau)$ into a subset of $\mathcal{V}(\theta)$ preserving the ancestor order, with an additional condition on the \underline{L}east \underline{C}ommon \underline{A}ncestors (LCAs) \cite[condition (2) p.\,208]{Zhang1996}: if, for $1\leq i\leq3$, $v_i\in\mathcal{V}(\tau)$ and $w_i=\varphi(v_i)\in\mathcal{V}(\theta)$, then $\LCA(v_1,v_2)$ is a proper ancestor of $v_3$ if and only if $\LCA(w_1,w_2)$ is a proper ancestor of $w_3$.

\medskip

\noindent
Let $\theta$ be a tree that approximates $\tau$ obtained by inserting nodes in $\tau$ only and consider the induced mapping $M_{\tau\to\theta}$ that associates nodes of $\tau$ with theirselves in $\theta$. We want the approximation process to be consistent with Zhang's edit distance $D_Z$, i.e., we want the mapping $M_{\tau\to\theta}$ to be a constrained mapping in the sense of Zhang, which in particular implies $D_Z(\theta,\tau)  = \#\mathcal{V}(\theta)-\#\mathcal{V}(\tau)$. We shall prove that this requirement excludes some inserting operations in our context.

\medskip

\begin{figure}
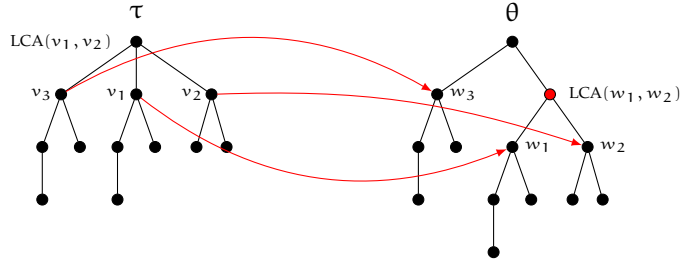

\centering\mappingA
\caption{The tree $\theta$ is obtained from $\tau$ by inserting an internal node. The associated mapping does not satisfy the conditions imposed by Zhang \cite{Zhang1996} because the LCA of $v_1$ and $v_2$ is a proper ancestor of $v_3$ whereas the LCA of $w_1$ and $w_2$ is not a proper ancestor of $w_3$.}
\label{fig:mappinginsert}
\end{figure}

\noindent
Indeed, the mapping $M_{\tau\to\theta}$ involved in the inserting operation of Fig.\,\ref{fig:insertnode} is partially displayed in Fig.\,\ref{fig:mappinginsert}, nodes $v_i$ of $\tau$ being associated with nodes $w_i$ of $\theta$. The LCA of $v_1$ and $v_2$ in $\tau$ is a proper ancestor of $v_3$. However, the LCA of $w_1$ and $w_2$ in $\theta$ is not a proper ancestor of $w_3$. As a consequence, this mapping is not a constrained mapping as defined by Zhang. A necessary and sufficient condition for $M_{\tau\to\theta}$ to be a constrained mapping is given in Lemma \ref{lem:cns:constrainedmapping}.

\begin{lemma}
\label{lem:cns:constrainedmapping}
Let $\tau$ be a tree and $v\in\mathcal{V}(\tau)$. Let $\theta$ be the tree obtained from $\tau$ by adding a node $w$ as a child of $v$ making the nodes of the subset $C\subset\mathcal{C}_\tau(v)$ children of $w$. The mapping $M_{\tau\to\theta}$ induced by these inserting operations is a constrained mapping in the sense of Zhang if and only if $C=\emptyset$, $\#C=1$ or $\#C=\#\mathcal{C}_\tau(v)$.
\end{lemma}

\noindent
\textit{Proof.} The proof is obvious if $v$ has one or two children. Thus we assume that $v$ has at least three children $c_1$, $c_2$ and $c_3$. In $\tau$, the LCA of $c_1$ and $c_2$ is $v$ and $v$ is an ancestor of $c_3$. Adding $w$ as the parent of $c_1$ and $c_2$ makes it the LCA of these two nodes, but not an ancestor of $c_3$ in $\theta$. The additional condition on the LCAs is then not satisfied. This problem appears only when making $w$ the parent of at least two children and of not all the children of $v$.\hfill{\color{OrangeRed}{\leafNE}}

\medskip

\begin{figure}[h]
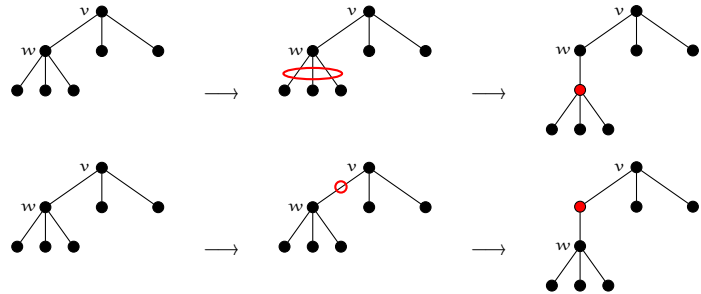

\centering\treeonechildallchildrenA\quad$\stackrel{\longrightarrow}{\text{\textcolor{white}{$\displaystyle\sum$}}}$\quad\treeonechildallchildrenB\quad$\stackrel{\longrightarrow}{\text{\textcolor{white}{$\displaystyle\sum$}}}$\quad\treeonechildallchildrenC\\
\treeonechildallchildrenA\quad$\stackrel{\longrightarrow}{\text{\textcolor{white}{$\displaystyle\sum$}}}$\quad\treeonechildallchildrenD\quad$\stackrel{\longrightarrow}{\text{\textcolor{white}{$\displaystyle\sum$}}}$\quad\treeonechildallchildrenE
\caption{Adding a node as new child of $w$ making all the current children of $w$ children of this new node (top) provides the same topology as adding a new node between $v$ and its child $w$ (bottom).}
\label{fig:onechildallchildren}
\end{figure}

\noindent
Consequently, we restrict ourselves to the following inserting operations which are the only ones that ensure that the associated mapping satisfies Zhang's condition: adding $w$ as a child of $v$ will make $w$ (i) a leaf, (ii) the parent of one current child of $v$, or (iii) the parent of all the current children of $v$. However, it should be noticed that (iii) can always be expressed as (ii) (see Fig.\,\ref{fig:onechildallchildren}). Finally, we only consider the inserting operations that make the new child of $v$ the parent of zero or one current child of $v$. For obvious reasons of symmetry, the allowed deleting operations are the complement of inserting operations, i.e., one can delete an internal node if and only if it has a unique child, which also ensures that the induced mapping is constrained in the sense of Zhang.

\subsubsection{Preserving the height of the pre-existing nodes}

In \cite[Definition 9 and Fig.\,6]{GF2010}, the NEST of a tree $\tau$ is obtained by successive partial linearizations of the (non-linear) DAG of $\tau$ which consist in merging all the nodes at the same height of the DAG. A consequence is that the height of any pre-existing node of $\tau$ is not changed by the inserting operations. For the sake of consistency with \cite{GF2010}, we only consider inserting and deleting operations that preserve the height of all the pre-existing nodes of $\tau$.

\medskip

\noindent
The next two results deal with inserting operations that preserve the height of the pre-existing nodes.

\begin{lemma}
\label{lem:insert:intern}
Let $\tau$ be a tree, $v\in\mathcal{V}(\tau)$ and $c\in\mathcal{C}_\tau(v)$. Let $\theta$ be the tree obtained from $\tau$ by adding the internal node $w$ as a child of $v$ making $w$ the parent of $c$. Then,
$$\forall\,u\in\mathcal{V}(\tau),~\mathcal{H}(\theta[u]) = \mathcal{H}(\tau[u])\quad\Longleftrightarrow\quad\mathcal{H}(\tau[c])+1<\mathcal{H}(\tau[v]).$$
\end{lemma}
\noindent
\textit{Proof.} Adding $w$ may only increase the height of $v$ and the one of its ancestors in $\tau$. If the height of $v$ is not changed by adding $w$, the height of its ancestors will not be modified. The height of $v$ remains unchanged if and only if the height of $w$ in $\theta$, i.e., $\mathcal{H}(\tau[c])+1$, is strictly less than the height of $\tau[v]$.\hfill{\color{OrangeRed}{\leafNE}}

\begin{lemma}
\label{lem:insert:subtree}
Let $\tau$ be a tree and $v\in\mathcal{V}(\tau)$. Let $\theta$ be the tree obtained from $\tau$ by adding a tree $t$ as a child of $v$. Then,
$$\forall\,u\in\mathcal{V}(\tau),~\mathcal{H}(\theta[u]) = \mathcal{H}(\tau[u])\quad\Longleftrightarrow\quad\mathcal{H}(t)+1\leq \mathcal{H}(\tau[v]).$$
\end{lemma}
\noindent
\textit{Proof.} Adding a subtree $t$ under $v$ may only increase the height of $v$ and the one of its ancestors in $\tau$. If the height of $v$ is not changed by adding $t$, the height of its ancestors will not be modified. Adding $t$ will make the height of $v$ increase if $\mathcal{H}(t)$ is strictly greater than the height of the higher child of $v$.\hfill{\color{OrangeRed}{\leafNE}}

\medskip

\noindent
A particular case of Lemma \ref{lem:insert:subtree} is the insertion of leaves in a tree. In light of the above result, a leaf can be added under $v$ if and only if $\mathcal{H}(\tau[v])\geq1$, i.e., $v$ is not a leaf. The below results concern deleting operations that preserve the height of the remaining nodes of $\tau$.

\begin{lemma}
\label{lem:del:intern}
Let $\tau$ be a tree, $v\in\mathcal{V}(\tau)$, $w\in\mathcal{C}_\tau(v)$ and $\mathcal{C}_\tau(w)=\{c\}$. Let $\theta$ be the tree obtained from $\tau$ by deleting the internal node $w$ making its unique child $c$ a child of $v$. Then,
$$\forall\,u\in\mathcal{V}(\theta),~\mathcal{H}(\theta[u])=\mathcal{H}(\tau[u])\quad\Longleftrightarrow\quad\exists\,w'\in\mathcal{C}_\tau(v)\setminus\{w\},~\mathcal{H}(\tau[w'])+1=\mathcal{H}(\tau[v]).$$
\end{lemma}

\noindent
\textit{Proof.} Deleting $w$ may only decrease the height of $v$ and the one of its ancestors in $\tau$. If the height of $v$ is not changed by deleting $w$, the height of its ancestors will not be modified. The height of $v$ remains unchanged if and only if it has a child different of $w$ of height $\mathcal{H}(\tau[v])-1$.\hfill{\color{OrangeRed}{\leafNE}}

\begin{lemma}
\label{lem:del:subtree}
Let $\tau$ be a tree, $v\in\mathcal{V}(\tau)$, $c\in\mathcal{C}_\tau(v)$. Let $\theta$ be the tree obtained from $\tau$ by deleting the subtree $\tau[c]$. Then,
$$\forall\,u\in\mathcal{V}(\theta),~\mathcal{H}(\theta[u])=\mathcal{H}(\tau[u])\quad\Longleftrightarrow\quad\exists\,c'\in\mathcal{C}_\tau(v)\setminus\{c\},~\mathcal{H}(\tau[c'])+1=\mathcal{H}(\tau[v]).$$
\end{lemma}

\noindent
\textit{Proof.} The proof follows the same reasoning as in the previous result.\hfill{\color{OrangeRed}{\leafNE}}

\subsubsection{NEST and NeST}

In view of the foregoing, we consider the set of inserting and deleting operations that fulfill the below requirements.

\medskip

\begin{figure}[h]
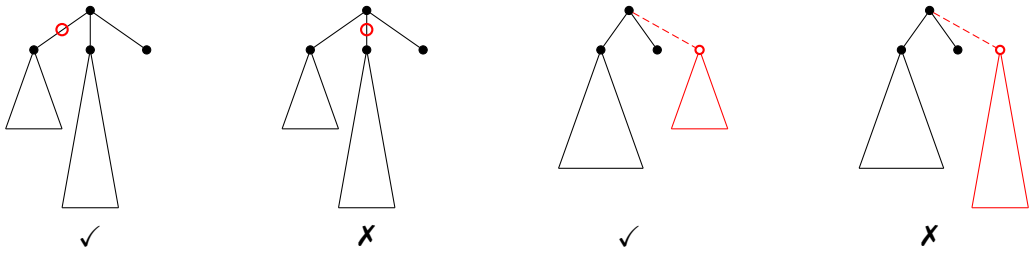

\centering
\operationsIB
\qquad\qquad
\operationsIC
\qquad\qquad
\operationsID
\qquad\qquad
\operationsIE
\caption{Allowed (\checkmark) and forbidden (\xmark) inserting operations to construct the NEST of a tree.}
\label{fig:insertions}
\end{figure}

\noindent
\textbf{\underline{A}dding operations} (see Fig.\,\ref{fig:insertions})
\begin{itemize}[leftmargin=18pt]
\item \underline{I}nternal nodes (AI): adding $w$ as a child of $v$ making $w$ the parent of the child $c$ of $v$ can be done only if $\mathcal{H}(\tau[c])+1<\mathcal{H}(\tau[v])$.
\item \underline{S}ubtrees (AS): adding $t$ as a child of $v$ can be done only if $\mathcal{H}(t)+1\leq\mathcal{H}(\tau[v])$.
\end{itemize}

\begin{figure}[h]
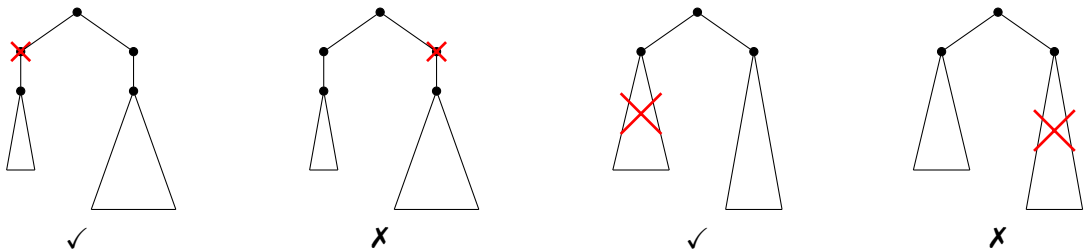

\centering
\operationsDA\qquad\qquad\operationsDB\qquad\qquad\operationsDC\qquad\qquad\operationsDD
\caption{Allowed (\checkmark) and forbidden (\xmark) deleting operations to construct the NeST of a tree.}
\label{fig:deletions}
\end{figure}

\noindent
\textbf{\underline{D}eleting operations} (see Fig.\,\ref{fig:deletions})
\begin{itemize}[leftmargin=18pt]
\item \underline{I}nternal nodes (DI): deleting $v\in\mathcal{C}_\tau(u)$ (making the unique child $w$ of $v$ a child of $u$) can be done only if there exists $v'\in\mathcal{C}_\tau(u)$, $v\neq v'$, such that $\mathcal{H}(\tau[v'])\geq\mathcal{H}(\tau[v])$.
\item \underline{S}ubtrees (DS): deleting the subtree $\tau[w]$, $w\in\mathcal{C}_\tau(v)$, of $\tau$ can be done if there exists $w'\in\mathcal{C}_\tau(v)$, $w'\neq w$, such that $\mathcal{H}(\tau[w'])+1=\mathcal{H}(\tau[v])$.
\end{itemize}

\begin{proposition}
The editing operations AI and AS (DI and DS, respectively) are the only inserting (deleting, respectively) operations that ensure that (i) the induced mapping is a constrained mapping and that (ii) the height of all the pre-existing nodes is unchanged.
\end{proposition}

\noindent
\textit{Proof.} This result is a direct corollary of Lemmas \ref{lem:cns:constrainedmapping}, \ref{lem:insert:intern}, \ref{lem:insert:subtree}, \ref{lem:del:intern} and \ref{lem:del:subtree}.\hfill{\color{OrangeRed}{\leafNE}}

\medskip

\noindent
The NEST (the NeST, respectively) of a tree $\tau$ is the self-nested tree obtained by the set of inserting operations AI and AS (of deleting operations DI and DS, respectively) of minimal cost, the cost of inserting a subtree being its number of nodes. Existence and uniqueness of the NEST are not obvious at this stage. The NeST exists because the (self-nested) tree composed of a unique root can be easily obtained by deleting operations from any tree, but its uniqueness is not evident.

\subsection{NEST algorithm}
\label{s:4:ss:NEST}

In order to present our NEST algorithm in a concise form in Algorithm \ref{algo:NEST}, we need to define the following operations involving two vectors $u$ and $v$ of the same size $n$ and a real number $\gamma$,
$$
\left\{
\begin{array}{ccccl}
u&\!+\!&v &=& (u_1+v_1\,,\,\dots\,,\,u_n+v_n),\\
u&\!+\!&\gamma & =& (u_1+\gamma\,,\,\dots\,,\,u_n +\gamma),\\
u&\!\vee\!&\gamma &=& (\max(u_1,\gamma)\,,\,\dots\,,\,\max(u_n,\gamma)).
\end{array}
\right.
$$
In other words, these operations must be understood component by component. In addition, in a condition, $u=0$ ($u\neq0$, respectively) means that for all $1\leq i\leq n$, $u_i=0$ ($u_i\neq0$, respectively). Finally, for $1\leq i\leq j\leq n$, $u_{i\dots j}$ denotes the vector $(u_i,\dots,u_j)$ of length $j-i+1$. This notation will also be used in Algorithm \ref{algo:NeST} for calculating the NeST. It should be noticed that an illustrative example that can help the reader to follow the progress of the algorithm is provided in Fig.\,\ref{fig:nestmax:kn} in Section~\ref{s:6}.

\SetKwFunction{NEST}{NEST}%
\SetKwProg{Fn}{Function}{\string:}{}
\begin{algorithm}
\DontPrintSemicolon
\Fn(){\NEST{$\tau$}}{
\KwData{the height profile $\rho$ of an unordered tree $\tau$}
\KwResult{the nearest embedding self-nested tree of $\tau$}

\For{$h_1$ {\bf{from}} $1$ {\bf{to}} $\mathcal{H}(\tau)$}{

	\For{$h_2$ {\bf{from}} $h_1-1$ {\bf{to}} $0$}{

		$\Delta\leftarrow\max\,\rho_{h_1,h_2} - \rho_{h_1,h_2}$\\
		$\rho_{h_1,h_2}\leftarrow\max\rho_{h_1,h_2}$\\
		$i\leftarrow1$
		
		\While{$\Delta\neq0$ {\bf{and}} $i\leq h_2$}{
			$\Delta~,~\rho_{h_1,h_2-i} \leftarrow(\Delta-\rho_{h_1,h_2-i})\vee0~,~\rho_{h_1,h_2-i}-\Delta$\\
			$i\leftarrow i+1$
		}
	}
}
\Return{\SN{$\rho$}}	
}
\caption{Construction of the nearest embedding self-nested tree}
\label{algo:NEST}
\end{algorithm}

\noindent
The relation between the above algorithm and the NEST of a tree is provided in the following result, which states in particular the existence of the NEST.

\begin{proposition}
\label{prop:res:NEST}
For any tree $\tau$, Algorithm \ref{algo:NEST} returns the unique NEST of $\tau$ in $O(\mathcal{H}(\tau)^2\times\mathcal{D}(\tau))$.
\end{proposition}
\noindent
\textit{Proof.} By definition of the NEST, the height of all the pre-existing nodes of $\tau$ can not be modified. Thus, the number of nodes of height $h-1$ under a node of height $h$ can only increase by inserting subtrees in the structure. Then we have
\begin{equation}
\label{eq:nest:1}
\rho_{\treeNEST(\tau)}(h,h-1)\geq\max\,\rho_\tau(h,h-1).
\end{equation}
Let $v$ be a vertex of height $h$ in $\tau$. We recall that $\gamma_i(v)$ denotes the number of subtrees of height $i$ under $v$. Our objective is to understand the consequences for $\gamma_i(v)$ of inserting operations to obtain $\rho_{\treeNEST(\tau)}(h,h-1)$ subtrees of height $h-1$ under $v$. To this aim, we shall define a sequence $\gamma_i^{(h-1,j)}(v)$ starting from $\gamma_i^{(h-1,0)}(v) = \gamma_i(v)$ that corresponds to the modified versions of $\tau$. The first exponent $h-1$ means that this sequence concerns editing operations used to get the good number of subtrees of height $h-1$ under $v$.

\smallskip

\noindent
Let $\Delta_{h-1}^{(0)}(v)=\rho_{\treeNEST(\tau)}(h,h-1)-\gamma_{h-1}^{(0)}(v)$ be the number of subtrees of height $h-1$ that must be added under $v$ to obtain the height profile of the NEST under $v$, i.e.,
$$\gamma_{h-1}^{(h-1,1)}(v) = \rho_{\treeNEST(\tau)}(h,h-1).$$
Implicitly, it means that $\gamma_{i}^{(h-1,1)}(v) = \gamma_i{(0)}(v)$ for $i\neq h-1$. The subtrees of height $h-1$ that we have to add are isomorphic, self-nested and embed all the subtrees of height $h-2$ appearing in $\tau$ by definition of the NEST. In particular, they can be obtained by the allowed inserting operations from the subtrees of height $h-2$ under $v$, by first adding an internal node to increase their height to $h-1$. In addition, it is less costly in terms of editing operations to construct the subtrees of height $h-1$ from the subtrees of height $h-2$ available under $v$ than to directly add these subtrees under $v$. If all the subtrees of height $h-2$ under $v$ must be reconstructed later, it will be possible to insert them and the total cost will be same as by directly adding the subtrees of height $h-1$ under $v$. As a consequence, all the available subtrees of height $h-2$ are used to construct subtrees of height $h-1$ under $v$ and it remains
$$ \Delta_{h-1}^{(1)} = \left(\Delta_{h-1}^{(0)}(v) - \gamma_{h-2}^{(h-1,1)}\right)\vee0$$
subtrees of height $h-1$ to be built under $v$. Furthermore, in the new version of $\tau$, we have
$$\gamma_{h-2}^{(h-1,2)}(v) = \gamma_{h-2}^{(h-1,1)}(v) - \Delta_{h-1}^{(1)}(v).$$
The $\Delta_{h-1}^{(1)}$ subtrees of height $h-1$ can be constructed from subtrees of height $h-3$ (with a larger cost than from subtrees of height $h-2$), and so on. To this aim, we define the sequence of the modified versions of $\tau$ by, for $0\leq j\leq h-2$,
$$
\left\{
\begin{array}{ccl}
\Delta_{h-1}^{(j+1)}(v) &=& \left(\Delta_{h-1}^{(j)}(v) - \gamma_{h-1-(j+1)}^{(h-1,j+1)}(v)\right)\vee0,\\
\gamma_{h-(j+2)}^{(h-1,j+2)}(v) &=& \gamma_{h-(j+2)}^{(h-1,j+1)}(v) - \Delta_{h-1}^{(j+1)}(v) .
\end{array}
\right.
$$
At the final step $j=h-2$, the $\Delta_{h-1}^{(0)}(v)$ subtrees of height $h-1$ have been constructed from all the available subtrees appearing under $v$, starting from subtrees of height $h-2$, then $h-3$, etc, and then have been added if necessary.

\smallskip

\noindent
From now on, the number of subtrees of height $h-2$ under $v$ will not decrease. Indeed, it would mean that an internal node has been added between $v$ and the root of a subtree of height $h-2$. This would have the consequence to increase of one unit the number of subtrees of height $h-1$ in subtrees of height $h$, which cost is (strictly) larger than adding a subtree of height $h-2$ in all the subtrees of height $h$. Consequently, we obtain
$$\rho_{\treeNEST(\tau)}(h,h-2) \geq \max_{\{v\in\mathcal{V}(\tau)\,:\,\mathcal{H}(\tau[v])=h\}} \gamma_{h-2}^{(h-1,h)}(v) .$$

\noindent
We can reproduce the above reasoning to construct under $v$ subtrees of height $h-i$, $i$ from $2$ to $h-1$, from subtrees with a smaller height, which defines a sequence $\gamma_{i}^{(h-i,j)}$ of modified versions of $\tau$, which size is $h-i+1$, and we get the following inequality,
\begin{equation}
\label{eq:nest:2}
\forall\,2\leq i\leq h,~\rho_{\treeNEST(\tau)}(h,h-i) \geq \max_{\{v\in\mathcal{V}(\tau)\,:\,\mathcal{H}(\tau[v])=h\}} \gamma_{h-i}^{(h-i+1,h-i+2)}(v)  .
\end{equation}

\noindent
The tree returned by Algorithm \ref{algo:NEST} is self-nested and its height profile saturates the inequalities \eqref{eq:nest:1} and \eqref{eq:nest:2} for all the possible values of $h$ and $i$ by construction. In addition, we have shown that this tree can be obtained from $\tau$ by the allowed inserting operations. Since increasing of one unit the height profile at $(h_1,h_2)$ has a (strictly) positive cost, this tree is thus the (unique) NEST of $\tau$. As seen previously, the number of iterations of the while loop at line 7 is the number of subtrees of height $h_2<h_1$ available to construct a tree of height $h_1$, i.e., the degree of $\tau$ in the worst case, which states the complexity.\hfill{\color{OrangeRed}{\leafNE}}

\subsection{NeST algorithm}
\label{s:4:ss:NeST}

This section is devoted to the presentation of the calculation of the NeST in Algorithm \ref{algo:NeST}. An illustrative example that can help the reader to follow the progress of the algorithm is provided in Fig.\,\ref{fig:nestmin:kn} in Section~\ref{s:6}.

\SetKwFunction{NeST}{NeST}%
\SetKwProg{Fn}{Function}{\string:}{}
\begin{algorithm}[h]
\DontPrintSemicolon
\Fn(){\NeST{$\tau$}}{
\KwData{the height profile $\rho$ of an unordered tree $\tau$}
\KwResult{the nearest embedded self-nested tree of $\tau$}

\For{$h_1$ {\bf{from}} $1$ {\bf{to}} $\mathcal{H}(\tau)$}{

	\For{$h_2$ {\bf{from}} $h_1-1$ {\bf{to}} $0$}{

		$\Delta\leftarrow\rho_{h_1,h_2} -\min\,\rho_{h_1,h_2}$\\
		$\rho_{h_1,h_2}\leftarrow\min\rho_{h_1,h_2}$
		
		\If{$\rho_{h_1-1,0\,\dots\,h_1-3}=0$~{\bf{and}}~$\rho_{h_1-1,h_1-2}=1$}{
			$\rho_{h_1,h_2-1}\leftarrow\rho_{h_1,h_2-1}+\Delta$
		}
	}
}
\Return{\SN{$\rho$}}	
}
\caption{Construction of the nearest embedded self-nested tree}
\label{algo:NeST}
\end{algorithm}

\begin{proposition}
\label{prop:res:NeST}
For any tree $\tau$, Algorithm \ref{algo:NeST} returns the unique NeST of $\tau$ in $O(\mathcal{H}(\tau)^2)$.
\end{proposition}
\noindent
\textit{Proof.} The proof follows the same reasoning as the proof of Proposition \ref{prop:res:NEST}. First, one may remark that
\begin{equation}
\label{eq:1:NeST}
\rho_{\treeNeST(\tau)}(h,h-1) \leq \min\,\rho_\tau(h,h-1),
\end{equation}
because the number of subtrees of height $h-1$ under a node $v$ of height $h$ can only decrease by the allowed deleting operations. Let $v$ be a node of height $h$ in $\tau$ and $\gamma_i(v)$ the number of subtrees of height $i$ under $v$. If a subtree of height $h-i$ under $v$ that has to be deleted is not self-nested, one can first modify it to get a self-nested tree and then remove it with the same overall cost. Thus, we can assume without loss of generality that all the subtrees under $v$ are self-nested. $\Delta_{h-1}(v)=\gamma_{h-1}(v)-\rho_{\treeNeST(\tau)}(h,h-1)$ denotes the number of subtrees of height $h-1$ that have to be removed from $v$. Let $\gamma_i^{(j)}(v)$ the sequence of the modifications to obtain $\rho_{\treeNeST(\tau)}(h,h-1)$ subtrees of height $h-1$ under $v$, with $\gamma_i^{(0)}(v) = \gamma_i(v)$. Instead of deleting a subtree of height $h-1$, it is always less costly to decrease its height of one unit by deleting its root. However it is possible only if this internal node has only one child, i.e., if $\rho_{\tau}(h-1,h-2)=1$ and $\rho_\tau(h-1,i)=0$ for $0\leq i<h-2$. If this new tree of height $h-2$ has to be deleted in the sequel, it will be done with the same global cost as by directly deleting the subtree of height $h-1$. As a consequence,
$$
\left\{
\begin{array}{ccl}
\gamma_{h-1}^{(1)}(v) &=&\rho_{\treeNeST(\tau)}(h,h-1),\\
\gamma_{h-2}^{(1)}(v) &=& \gamma_{h-2}^{(0)}(v) + \Delta_{h-1}(v)\mathbb{I}_{\{\rho_\tau(h-1,h-2)=1,\,\forall\,3\leq i\leq h,~\rho_\tau(h-1,h-i)=0\}}.
\end{array}
\right.
$$
From now on, the number of subtrees of height $h-2$ under $v$ will thus not increase and we obtain
$$\rho_{\treeNeST(\tau)}(h,h-2) \leq \min_{\{v\in\mathcal{V}(\tau)\,:\,\mathcal{H}(\tau[v])=h\}}\,\gamma^{(1)}_{h-2}(v).$$
There are $\Delta_{h-2}(v)=\gamma^{(1)}_{h-2}(v)-\rho_{\treeNeST(\tau)}(h,h-2)$ subtrees of height $h-2$ to be deleted under $v$. We can repeat the previous reasoning and delete the root of subtrees of height $h-2$ if possible rather than delete the whole structure, and so on for any height. Thus the sequence $\gamma_i^{(j)}$ is defined from
$$
\left\{
\begin{array}{ccl}
\Delta_{h-1-i}(v) &=&\gamma_{h-1-i}^{(i)}(v)- \rho_{\treeNeST(\tau)}(h,h-1-i), \\
\gamma_{h-1-i}^{(i+1)}(v) &=&  \rho_{\treeNeST(\tau)}(h,h-1-i),\\
\gamma_{h-2-i}^{(i+1)}(v) &=& \gamma_{h-2-i}^{(i)}(v) + \Delta_{h-1-i}(v)\mathbb{I}_{\{\rho_\tau(h-1,h-2)=1,\,\forall\,i+2\leq j\leq h,~\rho_\tau(h-i,h-j)=0\}},
\end{array}
\right.
$$
and we have
\begin{equation}
\label{eq:2:NeST}
\forall\,0\leq i\leq h-2,~\rho_{\treeNeST(\tau}(h,h-2-i) \leq \min_{\{v\in\mathcal{V}(\tau)\,:\,\mathcal{H}(\tau[v])=h\}}\, \gamma^{(i+1)}_{h-2-i}(v) .
\end{equation}
The tree returned by Algorithm \ref{algo:NeST} saturates the inequalities \eqref{eq:1:NeST} and \eqref{eq:2:NeST} for all the possible values of $h$ and $i$. Decreasing of one unit the height profile at $(h_1,h_2)$ has a (strictly) positive cost. Thus this tree is the (unique) NeST of $\tau$. The time-complexity is given by the size of the height profile array.\hfill{\color{OrangeRed}{\leafNE}}


\section{Numerical illustration}
\label{s:5}

\subsection{Random trees}
\label{ss:51}

\begin{figure}[h]
\centering
\includegraphics[width=7cm]{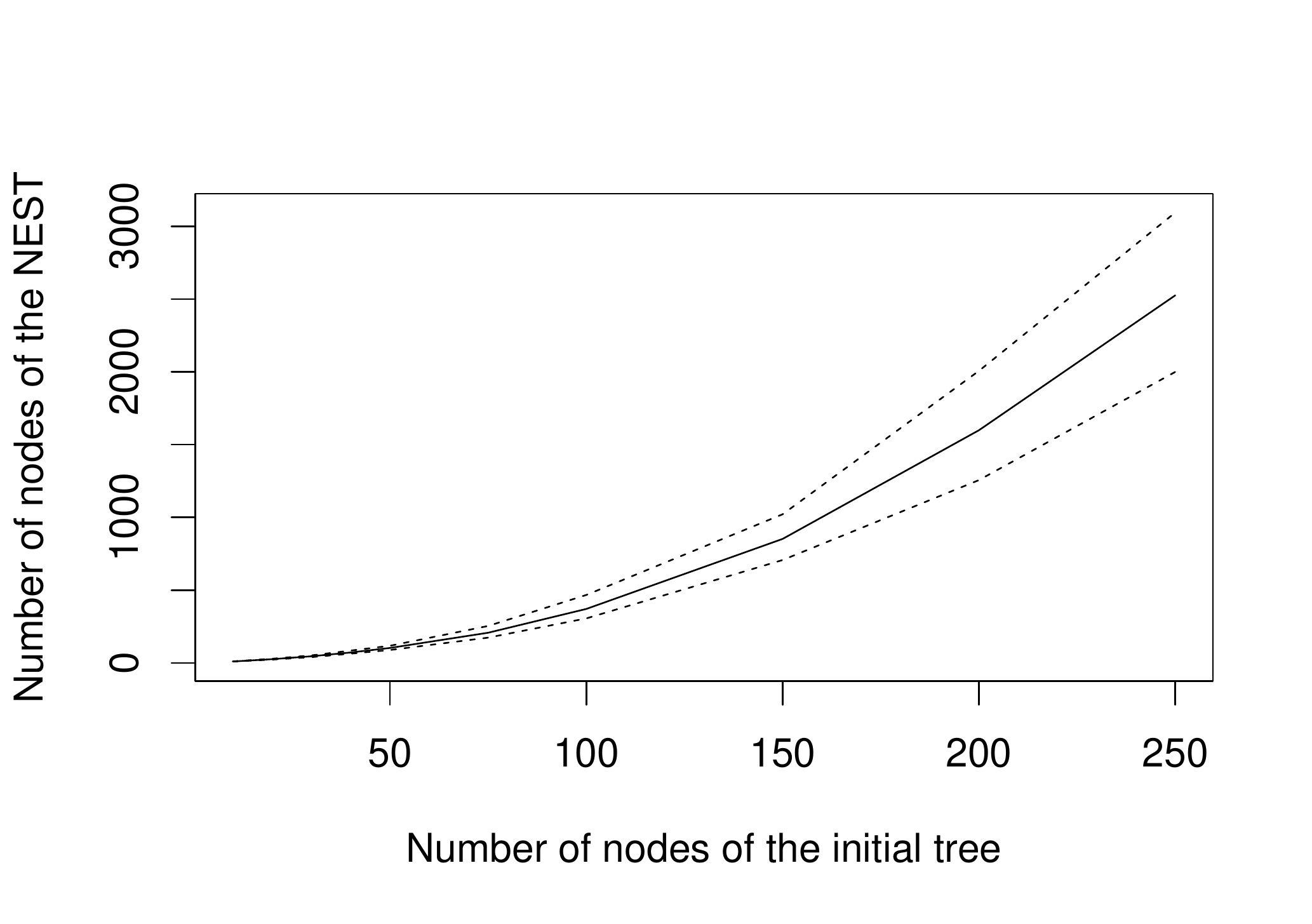}\quad\includegraphics[width=7cm]{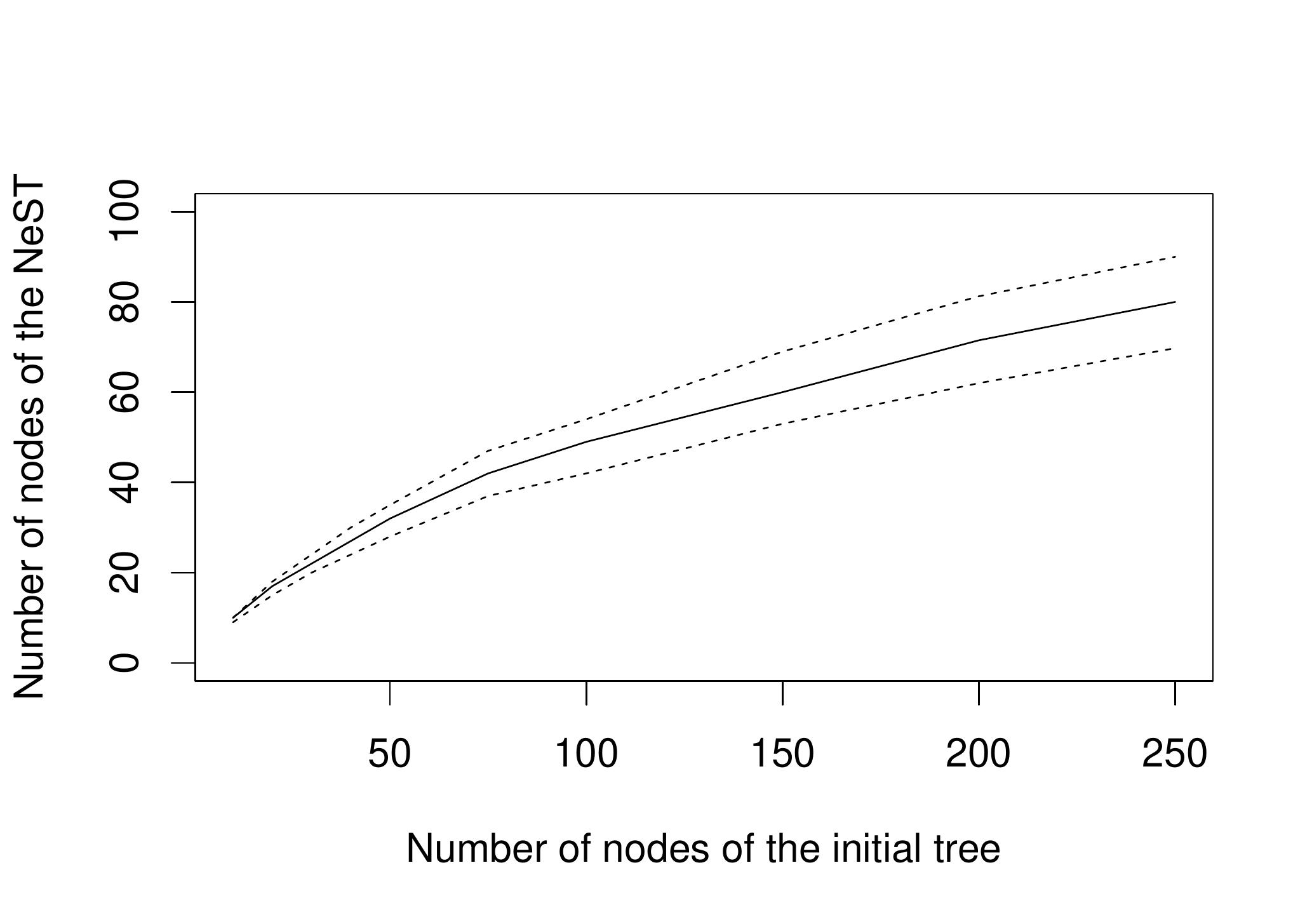}
\caption{Number of nodes of the NEST (left) and of the NeST (right) estimated from $3\,000$ random trees: average (full lines) and first and third quartiles (dashed lines).}
\label{fig:nest:nnodes}
\end{figure}

The aim of this section is to illustrate the behavior of the NEST and of the NeST on a set of simulated random trees regarding both the quality of the approximation and the computation time. We have simulated $3\,000$ random trees of size $10$, $20$, $30$, $40$, $50$, $75$, $100$, $150$, $200$ and $250$. For each tree, we have calculated the NEST and the NeST. The number of nodes of these approximations is displayed in Fig.\,\ref{fig:nest:nnodes}. We can observe that the number of nodes of the NEST is very large in regards with the size of the initial tree: approximately one thousand nodes on average for a tree of $150$ nodes, that is to say an approximation error of $750$ vertices. Remarkably, the NEST has never been a better approximation than the NeST on the set of simulated trees.

\medskip

\begin{figure}[h!]
\centering
\includegraphics[width=7cm]{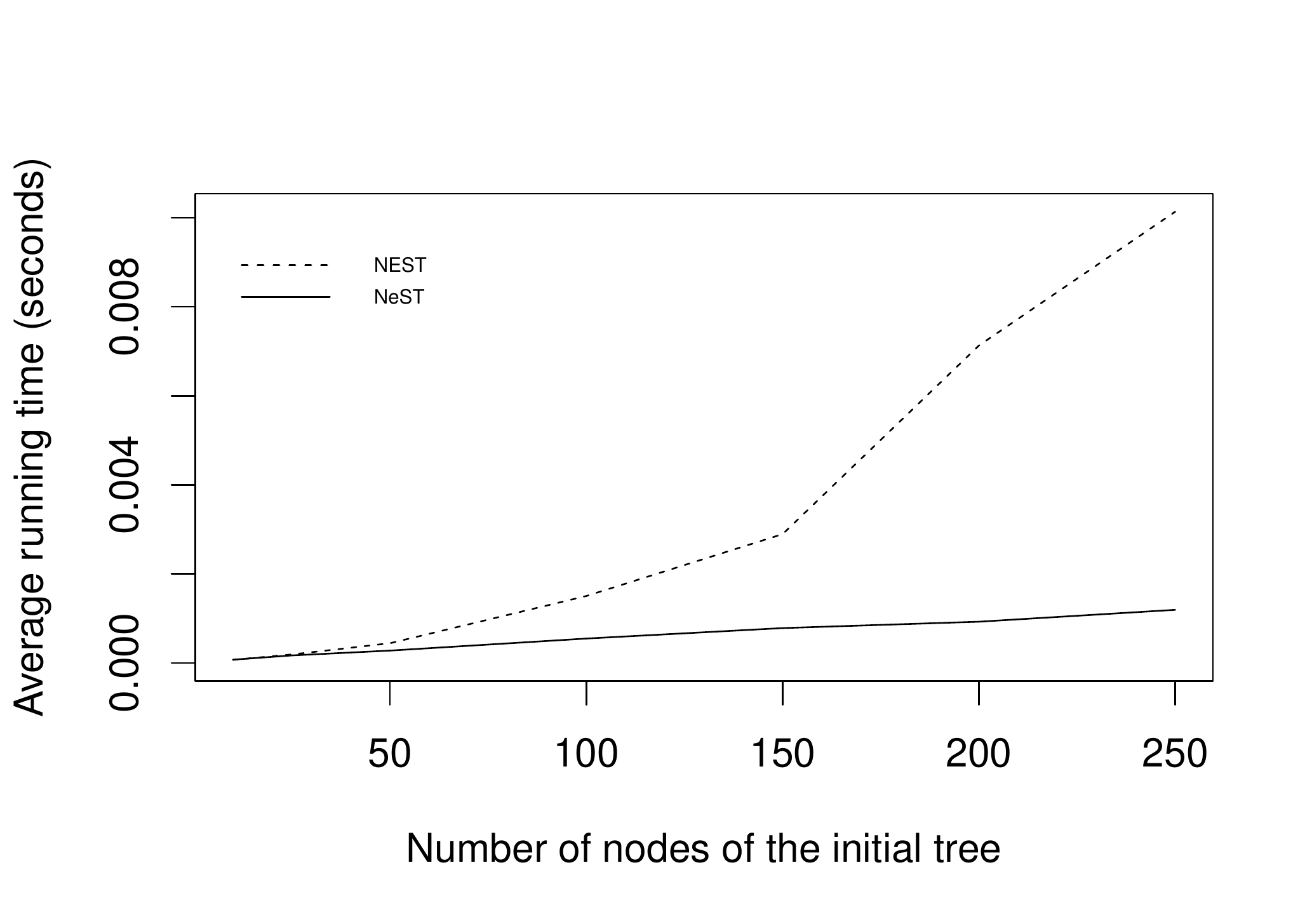}
\caption{Average running time required to compute the NEST (dashed line) or the NeST (full line) estimated from $3\,000$ simulated trees.}
\label{fig:time}
\end{figure}

\noindent
The computation time required to compute the NEST or the NeST of one tree on a 2.8 GHz Intel Core i7 has also been estimated on the set of simulated trees and is presented in Fig.\,\ref{fig:time}. As predicted by the theoretical complexities given in Propositions \ref{prop:res:NEST} and \ref{prop:res:NeST}, the NeST algorithm requires less computation time than the NEST. As a consequence, the NeST provides a much better and faster approximation of the initial data than the NEST.

\subsection{Structural analysis of a rice panicle}
\label{ss:52}

In light of \cite{GF2010}, we propose to quantify the degree of self-nestedness of a tree $\tau$ by the following indicator based on the calculation of $\treeNEST(\tau)$,
\begin{equation}
\delta_{\treeNEST}(\tau)  =   1-\frac{D_Z(\treeNEST(\tau),\tau)}{\#\mathcal{V}(\tau)}  =  \frac{2\#\mathcal{V}(\tau)-\#\mathcal{V}(\treeNEST(\tau))}{\#\mathcal{V}(\tau)}, \label{eq:def:deltaNEST}
\end{equation}
where $D_Z$ stands for Zhang's edit distance \cite{Zhang1996}. In \cite[eq.\,(6)]{GF2010}, the degree of self-nestedness of a plant is defined as in \eqref{eq:def:deltaNEST} but normalizing by the number of nodes of the NEST and not the size of the initial data, which avoids the indicator to be negative. In the present paper, we prefer normalizing by the number of nodes of $\tau$ to obtain the following comparable self-nestedness measure based on the calculation of $\treeNeST(\tau)$,
$$
\delta_{\treeNeST}(\tau) =1-\frac{D_Z(\treeNeST(\tau),\tau)}{\#\mathcal{V}(\tau)} =\frac{\#\mathcal{V}(\treeNeST(\tau))}{\#\mathcal{V}(\tau)} .
$$
The main advantage of this normalization is that, if the NEST and the NeST offer equally good approximations, i.e., $D_Z(\treeNEST(\tau),\tau)=D_Z(\treeNeST(\tau),\tau)$, then the degree of self-nestedness does not depend on the chosen approximation scheme, $\delta_{\treeNEST}(\tau)=\delta_{\treeNeST}(\tau)$.

\vspace{1cm}

\begin{figure}[h]
\centering
\includegraphics[width=7cm]{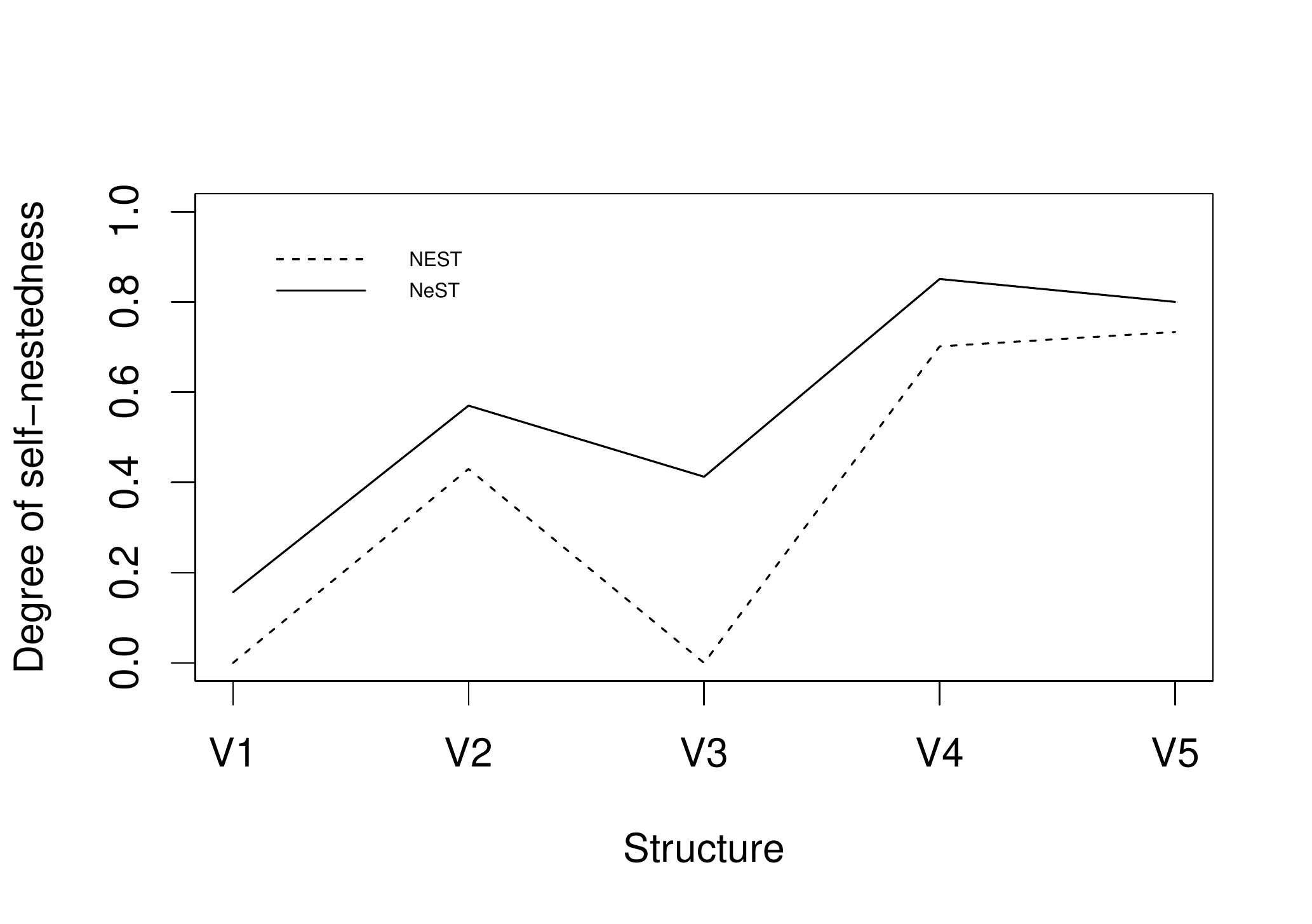}\quad\includegraphics[width=7cm]{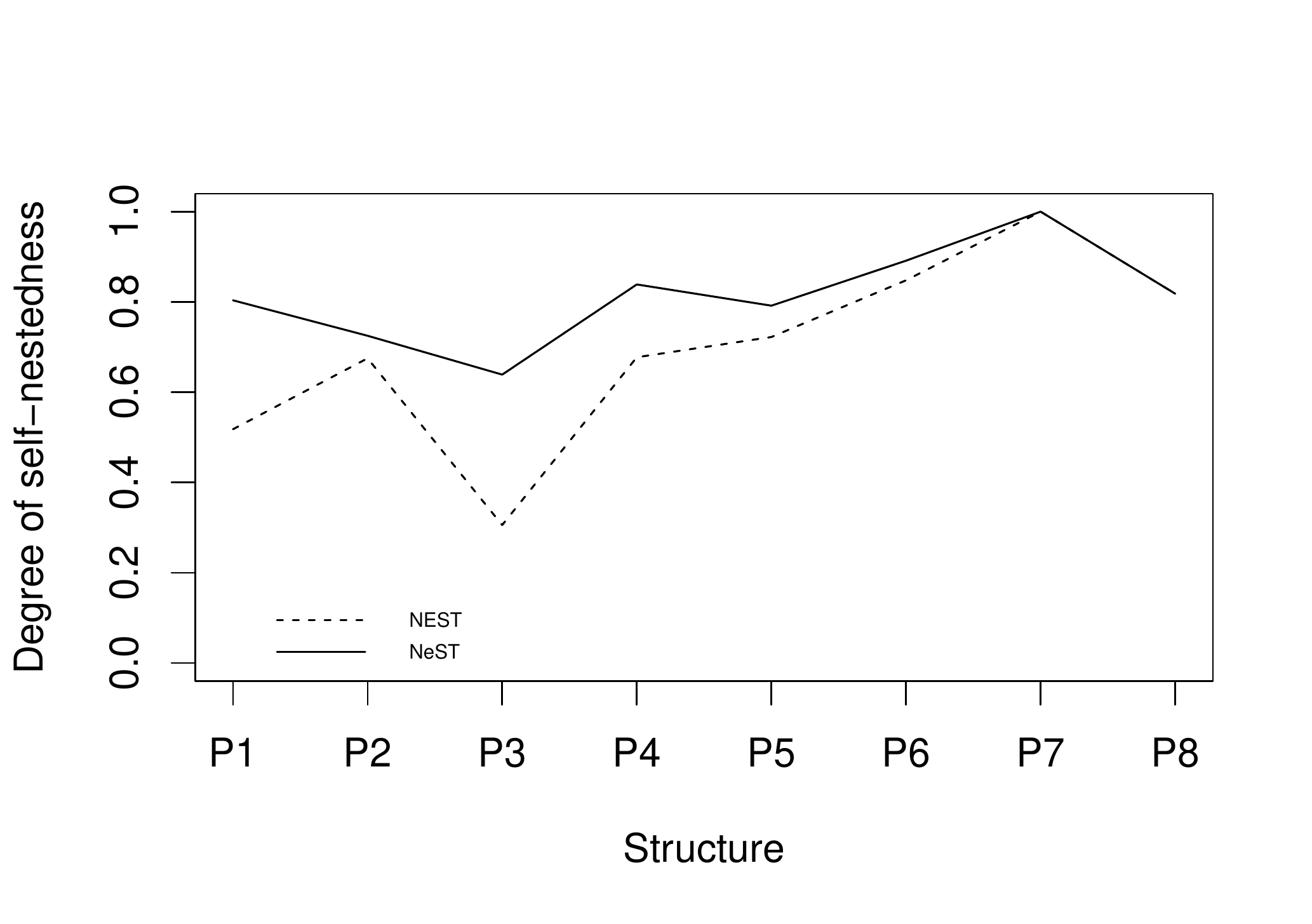}
\caption{Degree of self-nestedness measured by $\delta_{\treeNEST}\vee0$ (dashed lines) and $\delta_{\treeNeST}$ (full lines) of the different substructures appearing in the rice panicle.}
\label{fig:sn:panicle}
\end{figure}

\noindent
\begin{minipage}[l]{0.60\linewidth}
We propose to investigate the degree of structural self-similarity of the topological structure of the rice panicle studied in \cite[4.2 Analysis of a Real Plant]{GF2010} through these self-nested approximations. The rice panicle $V_1$ is made of a main axis bearing a main inflorescence $P_1$ and lateral systems $V_i$, $2\leq i\leq 5$, each composed of inflorescences $P_j$, $2\leq j\leq 8$ (see Fig.\,\ref{fig:rice}). We have computed the indicators of self-nestedness $\delta_{\treeNEST}\vee0$ and $\delta_{\treeNeST}$ for each substructure composing the whole panicle (see Fig.\,\ref{fig:sn:panicle}). The numerical values and the shape of these indicators are similar. However, $\delta_{\treeNeST}$ is always greater than $\delta_{\treeNEST}$, in particular for the largest structures $V_i$. Based on a better approximation procedure as highlighted in the previous section, the NeST better captures the self-nestedness of the rice panicle.
\end{minipage}\hfill
\begin{minipage}[r]{0.35\linewidth}
\centering
\includegraphics[width=4.15cm]{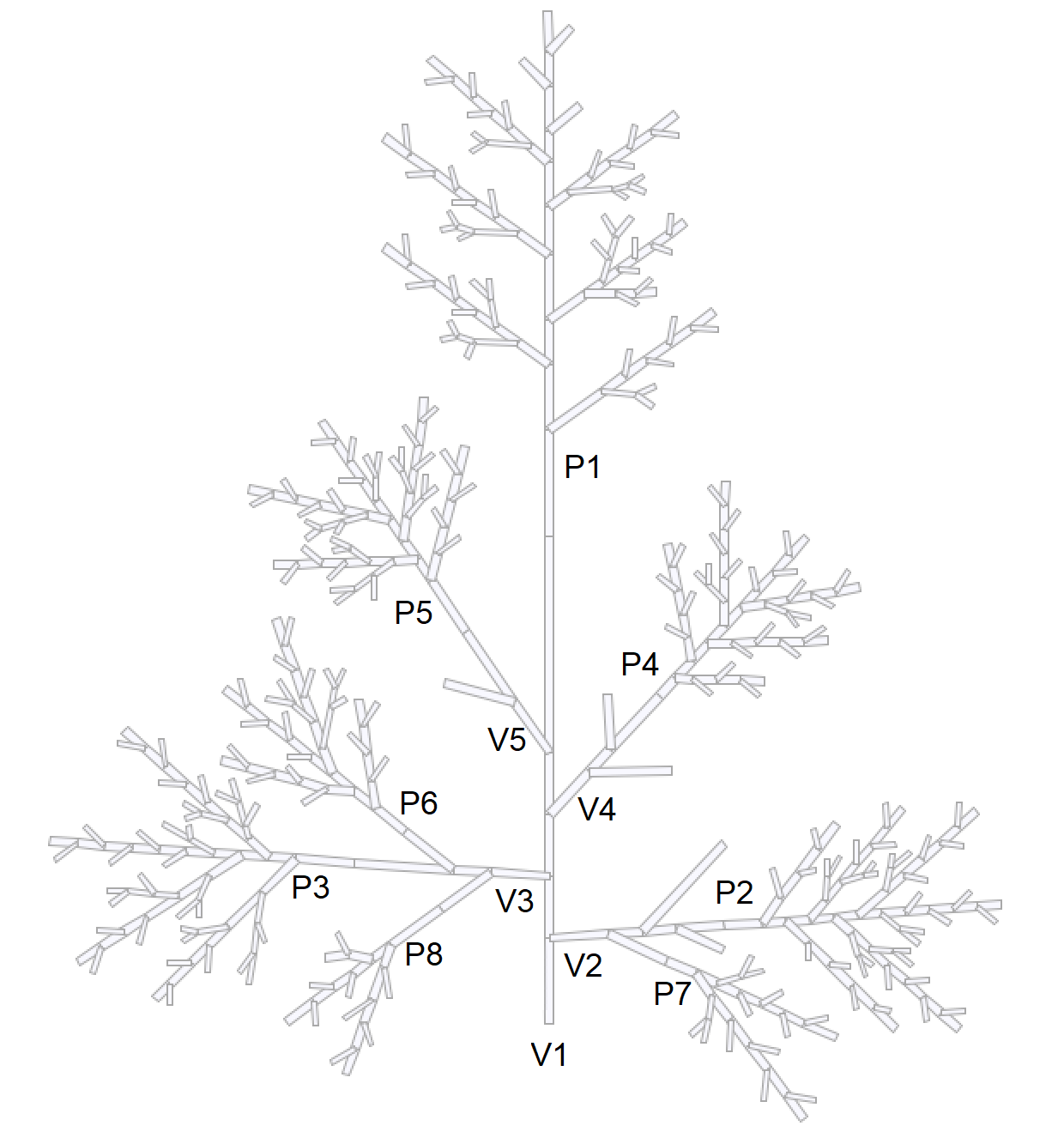}
\captionof{figure}{The rice panicle is composed of a main axis and lateral systems $V_i$, each made of one or several inflorescences $P_j$. \label{fig:rice}}
\end{minipage}

\section{Summary and concluding remarks}
\label{s:6}

Self-nested trees are unordered rooted trees that are the most compressed by DAG compression. Since DAG compression takes advantage of subtree repetitions, they present the highest level of redundancy in their subtrees.
In this paper, we have developed a new algorithm for computing the \underline{N}earest \underline{E}mbedding \underline{S}elf-nested \underline{T}ree (NEST) of a tree $\tau$ in $O(\mathcal{H}(\tau)^2\times\mathcal{D}(\tau))$, as well as the first algorithm for determining its \underline{N}earest \underline{e}mbedded \underline{S}elf-nested \underline{T}ree (NeST) with time-complexity $O(\mathcal{H}(\tau)^2)$.

\medskip

\begin{figure}[h]
\centering
\includegraphics[width=14cm]{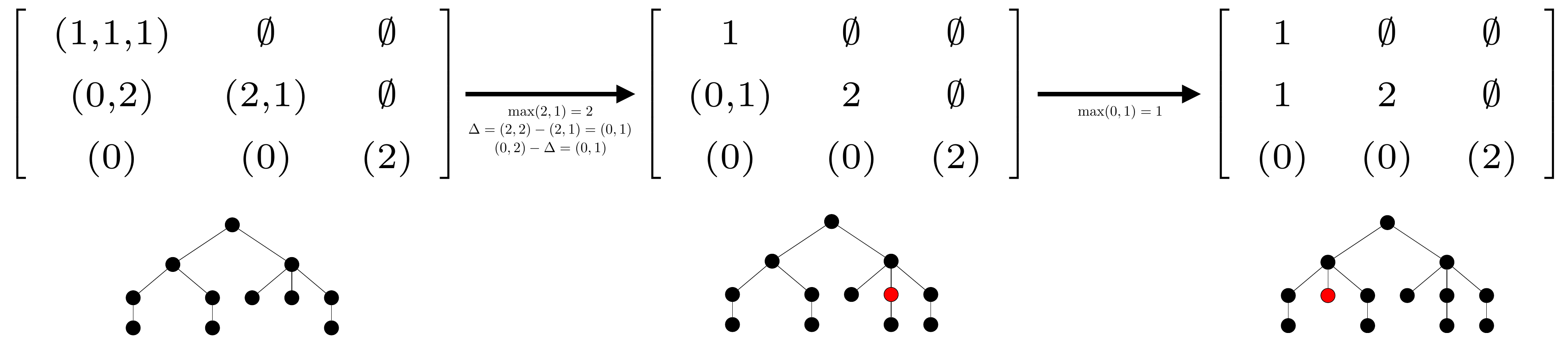}
\caption{Progress of Algorithm~\ref{algo:NEST} to compute the NEST of the left tree from its height profile. Only the second line has to be edited to get the correct output. Editions of the height profile are associated to addition of vertices in red. The output tree is self-nested and has been constructed by adding a minimal number of nodes to the initial tree.}
\label{fig:nestmax:kn}
\end{figure}

\begin{figure}[h]
\centering
\includegraphics[width=14cm]{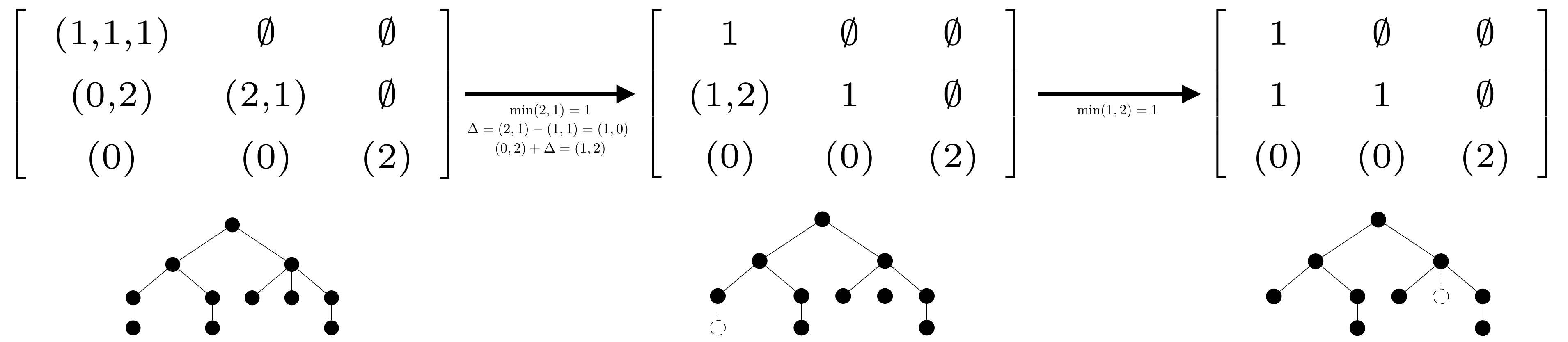}
\caption{Progress of Algorithm~\ref{algo:NeST} to compute the NeST of the left tree from its height profile. Only the second line has to be edited to get the correct output. Editions of the height profile are associated to deletion of vertices in dashed lines. The output tree is self-nested and has been constructed by removing a minimal number of nodes from the initial tree.}
\label{fig:nestmin:kn}
\end{figure}

\noindent
To this end, we have introduced the notion of height profile of a tree. Roughly speaking, the height profile is a triangular array which component $(h_1,h_2)$, with $h_2<h_1$, is the list of the numbers of direct subtrees of height $h_2$ in subtrees of height $h_1$, where a subtree is said direct if it is attached to the root.
We have shown in Proposition~\ref{prop:property:selfnested} that self-nested trees are characterized by their height profile.
While the first NEST algorithm \cite{GF2010} was based on edition of the DAG related to the tree to be compressed, the two approximation algorithms developed in the present paper take as input the height profile of any tree $\tau$, which can be computed in $O(\#\mathcal{V}(\tau)\times\mathcal{D}(\tau))$-time (see Proposition~\ref{prop:hp:complexity}), and modify it from top to bottom and from right to left, to return the self-nested height profile of the expected estimate (see Algorithms \ref{algo:NEST} and \ref{algo:NeST}). Figs.\,\ref{fig:nestmax:kn} and \ref{fig:nestmin:kn} illustrate the progress of the algorithms on a simple example. They should be examined in relation to the corresponding algorithms. We would like to emphasize that our paper also states the uniqueness of the NEST and of the NeST, and studies the link with edit operations admitted in Zhang's distance.

\noindent
Remarkably, estimations performed on a dataset of random trees establish that the NeST is a more accurate approximation of the initial tree than the NEST. This observation could be investigated from a theoretical perspective. In addition, we have shown that the NeST better captures the degree of structural self-similarity of a rice panicle than the NEST.

\medskip

\noindent
The algorithms developed in this paper are available in the last version of the \verb+Python+ library \verb+treex+ \cite{Azais2019treex}.

\section*{Acknowledgment}

The author would like to show his gratitude to two anonymous reviewers for their relevant comments on a first version of the manuscript.


\bibliography{main}

\end{document}